\documentclass[11pt, a4paper]{article}
\pdfoutput=1

\usepackage[utf8]{inputenc} 
\usepackage[T1]{fontenc}
\usepackage{amsmath}
\usepackage{amssymb}
\usepackage{amsfonts}
\usepackage{bbm}
\usepackage{verbatim}
\usepackage{dsfont}
\usepackage{booktabs}
\usepackage{slashed}
\usepackage{bm}
\usepackage{subcaption}
\usepackage{mathtools}
\usepackage{epsfig}
\usepackage{color}
\usepackage[table]{xcolor}
\usepackage{graphicx}
\usepackage[]{caption}
\usepackage{float}
\usepackage{overpic}
\usepackage{enumerate}
\usepackage{hhline}
\usepackage{multirow}
\usepackage{xspace}
\usepackage{setspace}
\usepackage[title]{appendix}
\usepackage{soul}
\usepackage{tensor}
\usepackage{tablefootnote}
\usepackage{diagbox}

\usepackage[square,numbers,compress]{natbib}
\usepackage{bibentry}

\usepackage[hidelinks, linkcolor=blue, citecolor=blue, urlcolor=blue]{hyperref}

\usepackage{tikz}
\usetikzlibrary{arrows}
\usetikzlibrary{calc}
\usetikzlibrary{decorations.pathreplacing,calligraphy}

\newcommand{\K}{{\cal K}}
\newcommand{\A}{{\cal A}}
\newcommand{\B}{{\cal B}}

\newcommand{\N}{{\cal N}}

\newcommand{\dd}{\text{d}}

\newcommand{\Z}{\mathbb{Z}}
\newcommand{\W}{{\cal W}}

\newcommand{\cR}{{\cal R}}

\renewcommand{\Re}{{\rm Re}\,}

\def\ov{\overline}
\def\im{\mbox{Im}\, }
\def\re{\mbox{Re}\, }
\def\om{\omega}

\def\IM{\text{Im}\,}
\def\RE{\text{Re}\,}
\def\CK {{\cal K}}
\newcommand{\cK}{\mathcal{K}}
\def\oh{\frac{1}{2}}

\renewcommand{\and}{\mbox{and}}

\usepackage{xcolor}

\newcommand{\blue}{\color{blue}}

\newcommand{\green}{\color{green!60!black}}

\newcommand{\MK}{M_{\rm KK}}
\newcommand{\MW}{M_{\rm W}}
\newcommand{\Mp}{M_{\rm p}}
\newcommand{\Rads}{R_{\rm AdS}}
\newcommand{\LUV}{\Lambda_{\rm UV}}

\newcommand{\half}{\frac{1}{2}} 

\renewcommand{\bm}{\boldmath} 

\def\be{\begin{equation}}
\def\ee{\end{equation}}
\def\bea{\begin{eqnarray}}
\def\eea{\end{eqnarray}}
\def\bes{\begin{subequations}}
\def\ees{\end{subequations}}

\allowdisplaybreaks

\counterwithin*{equation}{section}

\renewcommand{\baselinestretch}{1.5}

\begin{document}

\thispagestyle{empty}

\rightline{}

\begin{center}
\huge{{\bf Hiding the extra dimensions: A review on scale separation in string theory} 
\\[10mm]}
\large{Thibaut Coudarchet\footnote{coudarchet@thphys.uni-heidelberg.de}} \\[12mm]
\small{\it
Institute for Theoretical Physics, Heidelberg University,\\ Philosophenweg 19, 69120 Heidelberg, Germany 
\\[10mm]} 
\small{\bf Abstract} \\[5mm]
\end{center}
\begin{center}
\begin{minipage}[h]{15.0cm} 

We present an overview of both older and recent developments concerning scale separation in string theory. We focus on parametric scale separation obtained at the classical level in flux compactifications down to AdS vacua. We review the scenarios that have been proposed to achieve a hierarchy of scales between spacetime and the internal space, built from a low-dimensional perspective. We then discuss how they have been understood to arise from proper higher-dimensional descriptions. Eventually, limitations of these constructions as well as Swampland and holographic arguments addressing the question of scale separation in string theory are discussed. The purpose of the review is to draw an accurate picture of the state of the art of the subject at the moment.\\


\end{minipage}
\end{center}
\newpage

\setcounter{page}{1}
\pagestyle{plain}
\renewcommand{\thefootnote}{\arabic{footnote}}
\setcounter{footnote}{0}

\renewcommand{\baselinestretch}{1.5}


\tableofcontents

\section{Introduction}
\label{sec:Introduction}

Scale separation is the highly important phenomenological property for a given vacuum to have an internal compact space much ``smaller'' than the extended spacetime, so that a lower-dimensional effective description indeed makes sense. In this review, we focus on scale separation of Anti de Sitter (AdS) vacua where the scale associated to spacetime (the AdS radius) is set by the square root of the absolute value of the inverse cosmological constant. The decoupling or not of the extra dimensions is then measured by the Kaluza--Klein (KK) scale and how it compares with the AdS radius. We are also interested in a more stringent version of scale separation than the mere isolated property for a vacuum to feature a hierarchy of scales. Namely we investigate \emph{parametric scale separation} where we ask for the existence of a controlled infinite family of vacua showing stronger and stronger scale separation as one moves along the family. It is in the context of type IIA string theory and supergravity that most results concerning parametric scale separation have been achieved. In this framework, orientifold compactifications involving fluxes provide a rich playground and a variety of setups (in various dimensions, supersymmetric or not, with or without metric fluxes, etc) have been shown to give rise to infinite families of scale-separated vacua \cite{DeWolfe:2005uu,Narayan:2010em,Marchesano:2019hfb,Farakos:2020phe,VanHemelryck:2022ynr,Farakos:2023nms,Farakos:2023wps,Camara:2005dc,Derendinger:2004jn,Villadoro:2005cu,Marchesano:2020uqz,Cribiori:2021djm,Carrasco:2023hta}, in a controlled large volume and weak coupling regime.

The very interesting properties of such vacua legitimately motivated a large panel of works in the last twenty years to explore them further and in particular to understand their higher-dimensional origin to check their consistency. These constructions obtained from a low-dimensional effective perspective have been shown to be mapped to solutions of the 10$d$ equations of motion in the approximation where the fundamentally localised orientifold planes are actually smeared \cite{Acharya:2006ne} inside the whole internal space\footnote{Throughout the review we then use the expression ``low-dimensional effective perspective'', or any equivalent, to refer to the KK reduced theory inside which one can study the effect of diluted fluxes and where the O-planes are smeared.}. In this sense, the low-dimensional solutions only solve the zero mode of the 10$d$ equations, similarly to the KK truncation philosophy. This mapping between the (typically) 4$d$ and 10$d$ equations involving the smearing of sources have been debated/criticised a lot in the literature and still continues to be investigated. Indeed concerns have been raised \cite{Banks:2006hg,Douglas:2010rt,McOrist:2012yc} against the consistency of such a procedure to produce sensible solutions in the 4$d$ framework. For sure the solutions obtained in the smearing approximation cannot be seen in a trivial way to be good leading-order contributions of fully-fledged 10$d$ solutions, but a large variety of works showed that most concerns against the smearing could actually be overcome. Indeed it has been shown through different perspectives \cite{Cribiori:2021djm,Blaback:2010sj,Baines:2020dmu,Junghans:2020acz,Andriot:2023fss,Marchesano:2020qvg,Marchesano:2021ycx,Marchesano:2022rpr,Emelin:2021gzx} that true higher-dimensional solutions converge towards the smeared solutions almost everywhere in the internal space except at very short distances from the orientifold planes, where a stringy resolution thanks to F- or M-theory is expected to take place.

The existence of such M-theory embeddings for the scale-separated constructions is the next source of concern if one accepts that smearing is just fine. A large set of solutions including the original ones are obtained with a non-trivial Romans parameter whose presence complicates a proper quantisation of the string and an uplift to M-theory. Scale-separated solutions that can be seen as T-dual versions of the former have however been built \cite{Cribiori:2021djm,Carrasco:2023hta} which have interestingly a vanishing Romans mass, at the expense of introducing metric fluxes. Such solutions have been suggested in the toroidal case  \cite{Cribiori:2021djm} to be perfectly consistent both with the localisation of sources and with an M-theory uplift. These constructions then provide an optimistic answer to all concerns raised in the past towards the consistency of the scale-separated solutions.

New concerns however arose, in particular in light of the Swampland Program \cite{Vafa:2005ui,Brennan:2017rbf,Palti:2019pca,vanBeest:2021lhn,Grana:2021zvf} or, on the other side of the same coin, through the AdS/CFT correspondence \cite{Maldacena:1997re} and holographic arguments. Swampland statements act as no-go's for what can consistently be obtained at low energy from a quantum theory of gravity and it is natural to wonder how the property of scale separation is affected by these conjectures. Many constraints can be derived, in particular on the allowed amount of supersymmetry \cite{Cribiori:2022trc,Cribiori:2023ihv,Montero:2022ghl} thanks to the most well-established conjectures, while the most stringent of them like the AdS Distance Conjecture \cite{Lust:2019zwm} in its strong version seem to simply ban parametric scale separation in supersymmetric vacua from the Landscape, due to a fast decay of a tower of light states arising as the cosmological constant goes to zero. It remains a challenge to clearly support this strong version from basic principles but weaker versions and/or refinements \cite{Lust:2019zwm,Buratti:2020kda} arguing for softer decays of the light tower could be true and allow for scale separation. It is then clear that the status of scale separation is still up to debate and only future works will tell us where this nice property lies with respect to the Landscape borderline.

Similarly to the most severe conjectures, holographic arguments \cite{Heemskerk:2009pn,Polchinski:2009ch,deAlwis:2014wia,Conlon:2018vov,Conlon:2020wmc} cast a pessimistic shadow on scale-separated bulk spacetimes since properties of the would-be Conformal Field Theory (CFT) duals seem strange and, crucially, no dual is explicitly known so far. In the most famous AdS/CFT pairs like AdS$_5\times S^5$ in type IIB string theory or AdS$_4\times S^7$ in M-theory, the fluctuations on the spheres are large and comparable to the vacuum energy such that a low-dimensional description is inappropriate. A parametric separation of the scales associated with the two factors would imply a parametrically large gap in conformal dimensions between a few low-lying operators and an infinite number of others \cite{Polchinski:2009ch,deAlwis:2014wia,Conlon:2018vov,Conlon:2020wmc} and it is unclear if CFT's with such properties can exist. Moreover, after a specific analysis of holographic duals of gravitational theories with internal spaces that are Sasaki--Einstein manifolds coming from branes probing Calabi--Yau singularities, the authors of reference \cite{Collins:2022nux} conjectured that there is a universal upper bound on the lowest spin-2 conformal dimension that depends only on the number of dimensions of the internal space. This upper bound then translates into a lower bound on the diameter of the internal space expressed in AdS radius units which rules out parametric scale separation.

The purpose of this review is to provide a compact summary of the literature of roughly the last twenty years devoted to the problem of scale separation in string theory. We try to be as exhaustive as possible to review the various claims that have been made and how they have been addressed over the years though without going into too many technical details that would spoil the desired compactness of the review. The idea is thus to provide an accurate state of the art of scale separation in string theory with the hope to put light on problems and open questions that are worth investigating at the present day. Note that a nice review about scale separation can be found in \cite{talkVR} that was a source of inspiration to do this project, as well as in \cite{VanRiet:2023pnx}. As an important remark, notice also that we will often refer to some works and results as being ``criticisms'' of others but this is not at all to be taken in a pejorative way.

The rest of the review will follow what has been said in this introduction in chronological order: In sect.~\ref{sec:def} we start by defining what is meant by scale separation in an AdS framework. In sect.~\ref{sec:journey} after a recap of the type IIA effective theory machinery, we go through the various scenarios that have been proposed to achieve parametric scale separation, starting from the original DeWolfe, Giryavets, Kachru and Taylor (DGKT) proposal. We review its various generalisations/extensions that have been put forward and highlight their key ingredients. We then switch to the challenge of embedding these low-dimensional Effective Field Theory (EFT) constructions into fully-fledged 10$d$ supergravity theories in sect.~\ref{sec:higher}. We define the smeared approximation and review historical criticisms that have been made against it to then go through years of literature that have followed, which eventually justify the consistency of the approach. On a less optimistic perspective, we enter the Swampland realm in sect.~\ref{sec:conjs} to see how the net of swampland conjectures constrains scale separation and we discuss insights into the subject coming from holography. We finally end the review with a brief conclusion in sect.~\ref{sec:conclusion}.

\section{Separation of scales: Definitions}
\label{sec:def}

Considering string theory compactified on AdS$_d\times X_n$ with $n+d=10$, the would-be internal compact dimensions are characterised by two scales arising from the lowest-energy KK and winding modes, dictated by the geometry of the $n$-dimensional compact space $X_n$. For now we will use the volume $\text{Vol}_{X_n}$ of the internal space expressed in string units as a proxy to evaluate the KK and winding scales. Of course in reality the internal manifold can be largely anisotropic and expressing correctly the scales requires precise knowledge of the homology structure as well as the size of the individual cycles. The correct expressions for the main case of interest in this review, namely when $X_n$ is a Calabi--Yau threefold, can be found in sect.~\ref{sec:vanilla}. Using the volume proxy, the KK and winding scales $\MK$ and $\MW$ respectively read
\begin{equation}
\MK\equiv\frac{M_{\rm s}}{\text{Vol}_{X_n}^{1/n}}\, ,\qquad\qquad \MW\equiv M_{\rm s}{\text{Vol}_{X_n}^{1/n}}\, ,
\end{equation}
where $M_{\rm s}$ is the string mass scale defined with the following conventions: $M_{\rm s}\equiv\ell_{\rm s}^{-1}\equiv (2\pi\sqrt{\alpha'})^{-1}$. The $d$-dimensional dilaton $\phi_d$ relates the string scale to the reduced $d$-dimensional Planck scale\footnote{The reduced Planck scale is related to the usual one $\tilde\Mp$ with $8\pi\Mp^{d-2}=\tilde\Mp^{d-2}$ \cite{Polchinski:1998rq,Polchinski:2010hw}. Note that we do not add a subscript $d$ to indicate that the Planck mass is $d$-dimensional. It will always be understood in the following that $\Mp$ stands for the Planck mass in the appropriate number of extended dimensions, which most of time will be four.} $\Mp$ like $M_{\rm s}=(4\pi)^{-1/(d-2)}e^{\phi_d}\Mp$ and is defined as
\begin{equation}
\phi_d\equiv\frac{2}{d-2}\phi-\frac{1}{d-2}\log\text{Vol}_{X_n}\, ,
\end{equation}
where $\phi$ is the ten-dimensional dilaton (with $g_{\rm s}\equiv e^\phi$). Forgetting the $\pi$ factors which are irrelevant for our discussion, we can thus write
\begin{equation}
\label{eq:KKW}
\frac{\MK}{\Mp}\propto\frac{e^{\phi_d}}{\text{Vol}_{X_n}^{1/n}}\, ,\qquad\qquad \frac{\MW}{\Mp}\propto e^{\phi_d}\text{Vol}_{X_n}^{1/n}\, .
\end{equation}

On the other hand, the $d$-dimensional non-compact external AdS spacetime is characterised by a cosmological constant $\Lambda$ that we will express in $d$-dimensional Planck units. To be more precise, the $d$-dimensional action $S_d$ in Einstein frame reads
\begin{equation}
S_d=\frac{1}{\kappa_d^2}\int\dd^{d}x\sqrt{g}\left(\frac{\mathcal{R}}{2}-\kappa_d^2 V+\cdots\right)\, ,
\end{equation}
where $\kappa_d^2\equiv 8\pi G_d\equiv 1/\Mp^{d-2}$ and $V$ is the (dimensionful) scalar potential. The cosmological constant expressed in Planck units is then $\Lambda\equiv V/\Mp^d$. It defines what is called the \emph{AdS radius}, $\Rads$, or \emph{Hubble length}, expressed as \cite{Palti:2019pca,Nastase:2015wjb}
\begin{equation}
\label{eq:Rads}
(\Rads\Mp)^{-1}\equiv\frac{\sqrt{2}\sqrt{|\Lambda|}}{\sqrt{(d-1)(d-2)}}\sim \sqrt{|\Lambda|}\, ,
\end{equation}
where in the last approximate equality we forget about the $d$-dependent prefactor. Note that this prefactor is irrelevant when discussing how the AdS radius scales but will be important for the discussion of sect.~\ref{sec:integer}.

Now the whole point of scale separation is to look at how these three scales introduced so far compare to each other at a vacuum. We say that a vacuum is scale-separated if the KK and winding mass scales are much bigger than the inverse AdS radius, i.e when
\begin{equation}
\label{eq:scale_sep}
\text{Scale separation condition: }\quad \Rads\MK\gg 1 \quad\text{ and }\quad \Rads\MW\gg 1\, .
\end{equation}
The intuition behind this separation of scale is easy to grasp: It is only at scale-separated vacua that one can properly talk about a lower-dimensional effective theory, KK reduced and truncated, with its non-compact spacetime decoupled from a small compact internal space. On the contrary, if the scales have the same magnitude, this means that it makes little sense to discriminate some dimensions and call them ``internal'' in contrast to ``external'' ones. In this case, the theory needs to be understood as an intrinsically higher-dimensional one. Of course our universe has a clear four-dimensional external spacetime and that is why scale-separation is an important concept: A phenomenologically realistic vacuum has to be (in particular) scale separated. Notice that an AdS vacuum comes with another length scale $L_\rho$ \cite{Cribiori:2021djm} related to the cosmological constant scale like $(L_\rho\Mp)^{-1}=|\Lambda|^{\frac{1}{d}}$. Asking for separation of this scale from the KK scale is a much stronger requirement and this is not the hierarchy we will focus on in the following. 

One often cares about not only being able to achieve scale separation at a specific vacuum but about the possibility to generate an infinite family (typically by scaling fluxes) of such vacua. One then talks about \emph{parametric scale separation} as the separation of scales becomes bigger and bigger as one moves along the family by varying some parameters. Of course by doing so one must also ensure that the family of vacua stays in a controlled regime where the EFT description makes sense. Notice that this parametric definition rules out KKLT- \cite{Kachru:2003aw} and LVS-like \cite{Balasubramanian:2005zx} scenarios which are scale separated but only because of finely tuned quanta which result in isolated solutions found with the help of quantum corrections. We do not claim that anything is wrong with these scenarios before uplift to de Sitter (nor after because it is not the subject of this review) and that they do not have scale separation, but we will only focus on parametric scale separation in the following and mention KKLT only very quickly. We also focus on scale separation obtained at the classical level but it is worth mentioning that M-theory solutions featuring parametric scale separation thanks to Casimir energy corrections have been built in \cite{Luca:2022inb}, that we will briefly mention again in sect.~\ref{sec:casimir}.

It has been noted \cite{talkVR,VanRiet:2023pnx,Danielsson:2018ztv,talkmiguel} that the problem of achieving scale separation in string theory is very similar to the issues related to the cosmological constant problem. Indeed the subtleties arising when the extra dimensions decouple to leave a lower-dimensional theory match the ones that plague de Sitter (dS) constructions. One simple way to realise this connection is that the conditions for scale separation \eqref{eq:scale_sep} actually amount to ask for a decoupling between the KK scale and the cosmological constant thanks to \eqref{eq:Rads}. Then if the KK scale is a cutoff for an EFT, one naively expects a vacuum energy to be of the same order and no hierarchy appears. This lack of hierarchy then precisely reflects the absence of scale separation.

\section{A journey through proposed scale-separated vacua}
\label{sec:journey}

We can now make a tour of the different constructions that exist in the literature and that achieve parametric scale separation. As mentioned in the introduction, these vacua have been derived mostly from an effective 4$d$ perspective, leaving the trustworthiness of the solutions and the question of the existence of a fully-fledged 10$d$ uplift\footnote{Here the term uplift is used to designate a 10$d$ (or 11$d$) understanding of a low-dimensional, KK truncated, construction. Uplift in the sense of turning an AdS spacetime into dS is not the concern of this review (see \cite{Bena:2023sks} for an ironic state of the art on that subject).} unanswered. A lot of works have been dedicated to filling these caveats by unravelling and understanding better the 10$d$ completion of the aforementioned solutions, from the smearing approximation (see sect.~\ref{sec:smeared}) towards localisation of sources and backreaction control (see sect.~\ref{sec:smearing?}). As already mentioned, note that the questions concerning the 10$d$ uplift are not the only shadows lingering above scale-separated constructions but these other considerations will be reviewed later in sect.~\ref{sec:conjs}. Before delving into all this, it is useful to provide a visual summary of what has been explored in the literature to achieve scale separation and that will be reviewed in this section.

\subsection{A map of scale separation}

The following figure \ref{fig:map} summarises the various constructions found in the literature to achieve parametric scale separation at the classical level, as well as the directions followed to understand their higher-dimensional origins. Note that this map kind of only displays the optimistic results but the concerns regarding these constructions will of course be tackled in the review. The figure is actually the road-map of the present section and the following one, and we begin with a review of the proposed scale-separated vacua built from a low-dimensional perspective.

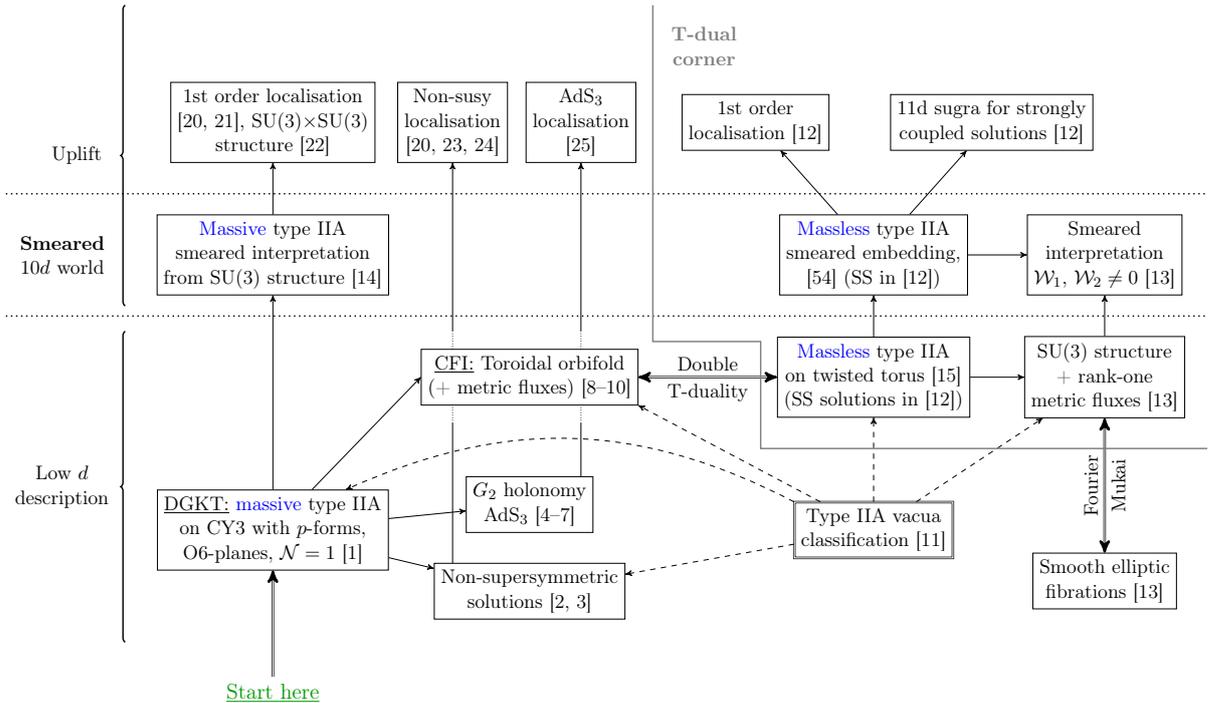
\begin{figure}[!ht]
\resizebox{\textwidth}{!}{
\begin{tikzpicture} [>=stealth',auto,node distance=3cm,main node/.style={circle,draw,font=\sffamily\Large\bfseries},scale=1, every node/.style={transform shape}]
  
\def\xone{-5.8}
\def\xtwo{-0.8}
\def\xthree{5.9}
\def\xfour{10.4}
\def\xfive{13.4}
\def\yone{-3}
\def\ytwo{0.2}
\def\ythree{3.2}
\def\yfour{5.6}
\def\yfive{8.2}
\def\ysix{10}
\def\xleg{2}
\def\yleg{-3}

\node[align=center] (start) at (\xone,\yone) {\green\large \underline{Start here}};
\node[draw, align=center] (DGKT) at (\xone,\ytwo) {\underline{DGKT:} {\blue massive} type IIA\\ on CY3 with $p$-forms,\\ O$6$-planes, $\N=1$ \cite{DeWolfe:2005uu}};
\node[draw, align=center] (N0) at (\xtwo,\ytwo-1.2) {Non-supersymmetric\\ solutions \cite{Narayan:2010em,Marchesano:2019hfb}};
\coordinate (N0left) at (\xtwo-1.5,\ytwo-0.65);
\node[draw, align=center] (G2) at (\xtwo,\ytwo+0.5) {$G_2$ holonomy\\ AdS$_3$ \cite{Farakos:2020phe,VanHemelryck:2022ynr,Farakos:2023nms,Farakos:2023wps}};
\coordinate (G2right) at (\xtwo+1,\ytwo+1.05);
\coordinate (CFIdown) at (\xtwo+1,\ythree-0.9);
\coordinate (CFIdownright) at (\xtwo+1,\ythree-0.55);
\coordinate (CFIdown2) at (\xtwo-1.5,\ythree-0.9);
\coordinate (CFIdownleft) at (\xtwo-1.5,\ythree-0.55);
\node[draw, align=center] (CFI) at (\xtwo,\ythree) {\underline{CFI:} Toroidal orbifold\\ (+ metric fluxes) \cite{Camara:2005dc,Derendinger:2004jn,Villadoro:2005cu}};
\coordinate (CFIup) at (\xtwo+1,\ythree+0.9);
\coordinate (CFIupright) at (\xtwo+1,\ythree+0.55);
\coordinate (CFIup2) at (\xtwo-1.5,\ythree+0.9);
\coordinate (CFIupleft) at (\xtwo-1.5,\ythree+0.55);
\node[draw,double,double distance=0.6pt,align=center] (syst) at (\xthree,\ytwo) {Type IIA vacua\\ classification \cite{Marchesano:2020uqz}};
\node[draw, align=center] (smeared) at (\xone,\yfour) {{\blue Massive} type IIA\\ smeared interpretation\\ from SU(3) structure \cite{Acharya:2006ne}};
\node[draw, align=center] (1st) at (\xone,\yfive) {1st order localisation\\\cite{Junghans:2020acz,Andriot:2023fss}, SU(3)$\times$SU(3)\\ structure \cite{Marchesano:2020qvg}};
\node[draw, align=center] (1st3d) at (\xtwo+1,\yfive) {AdS$_3$\\ localisation\\ \cite{Emelin:2021gzx}};
\node[draw, align=center] (1stN0) at (\xtwo-1.5,\yfive) {Non-susy\\ localisation\\
\cite{Junghans:2020acz,Marchesano:2021ycx,Marchesano:2022rpr}};
\node[draw, align=center] (TCFI) at (\xthree,\ythree) {{\blue Massless} type IIA\\ on twisted torus \cite{Banks:2006hg}\\(SS solutions in \cite{Cribiori:2021djm})};
\node[draw, align=center] (TCFIsmeared) at (\xthree,\yfour) {{\blue Massless} type IIA\\smeared embedding,\\\cite{Caviezel:2008ik} (SS in \cite{Cribiori:2021djm})};
\node[draw, align=center] (1stbis) at (\xthree-2.3,\yfive) {1st order \\ localisation  \cite{Cribiori:2021djm}};
\node[draw, align=center] (sugra) at (\xthree+2.3,\yfive) {$11$d sugra for strongly\\ coupled solutions  \cite{Cribiori:2021djm}};
\node[draw,align=center] (Tus) at (\xfour,\ytwo-1) {Smooth elliptic\\fibrations \cite{Carrasco:2023hta}};
\node[draw,align=center] (us) at (\xfour,\ythree) {SU(3) structure\\ + rank-one\\ metric fluxes \cite{Carrasco:2023hta}};
\node[draw,align=center] (upus) at (\xfour,\yfour) {Smeared\\ interpretation\\ $\W_1$, $\W_2\neq 0$ \cite{Carrasco:2023hta}};
\draw [->,double] (start) -- (DGKT);
\draw [->] (DGKT) --  (N0);
\draw [->] (DGKT) -- (G2);
\draw [->] (DGKT) -- (CFI.west);
\draw [->,dashed] (syst) -- node[above,rotate=11] {} (N0);
\draw [->,dashed] (syst) -- (CFI.south east);
\draw [->,dashed] (syst.north west) to [bend right=26] node[above,rotate=18,very near end] {} (DGKT);
\draw [->] (DGKT) -- (smeared);
\draw [->] (smeared) -- (1st);
\draw [-] (G2right) -- (CFIdown);
\draw [-,densely dotted] (CFIdown) -- (CFIdownright);
\draw [-,densely dotted] (CFIupright) -- (CFIup);
\draw [->] (CFIup) -- (1st3d);
\draw [-] (N0left) -- (CFIdown2);
\draw [-,densely dotted] (CFIdown2) -- (CFIdownleft);
\draw [-,densely dotted] (CFIupleft) -- (CFIup2);
\draw [->] (CFIup2) -- (1stN0);
\draw [<->,double] (CFI) -- node[above] {Double} node[below] {T-duality} (TCFI);
\draw [->] (TCFI) -- (TCFIsmeared);
\draw [->] (TCFIsmeared) -- (1stbis);
\draw [->] (TCFIsmeared) -- (sugra);
\draw [->] (TCFIsmeared) -- (upus);
\draw [<->,double] (Tus) -- node[above,rotate=90] {Fourier} node[below,rotate=90] {Mukai} (us);
\draw [->,dashed] (syst) -- (us);
\draw [->,dashed] (syst) -- (TCFI);
\draw [->] (TCFI) -- (us);
\draw [->] (us) -- (upus);
\draw[decoration={brace,raise=5pt},decorate,thick,align=center] (-8.5,-2) -- node[left=10pt] {Low $d$\\ description} (-8.5,4.1);
\draw[decoration={brace,raise=5pt},decorate,thick] (-8.5,4.6) -- node[left=14pt] {Uplift} (-8.5,10.5);
\node[align=center] (smeared) at (-9.9,5.6) {\textbf{Smeared}\\10$d$
 world};
\draw[dotted, thick] (-11,4.4) -- (12.4,4.4);
\draw[dotted, thick] (-11,6.8) -- (12.4,6.8);
\draw[gray, thick] (1.6,10.5) -- (1.6,3.9) -- (3.7,3.9) -- (3.7,1.8) -- (12.4,1.8);
\node[align=center,gray] (Tdual) at (2.6,9.7) {\textbf{T-dual}\\ \textbf{corner}};
\end{tikzpicture}
}
\caption{A map of classical and parametric scale separation in AdS vacua. Solid arrows denote generalisations or extensions from one construction to another while dashed arrows indicate subcases. The three regions delimited by the dotted lines constitutes the outline of this section and the following one. We first review the scenarios proposed to achieve scale separation built from a low-dimensional perspective. We then review how these constructions are understood as ten-dimensional solutions with a smeared approximation. Eventually, we review the concerns that have been raised against the consistency of smearing and we discuss the works going beyond this approximation that tackle the problem of localisation of the sources, their backreaction and the question of an M-theory embedding. Note that SS stands for Scale Separation.}
\label{fig:map}
\end{figure}

\subsection{Rudiments of type IIA flux compactifications}
\label{sec:4d}

The bottom-up development of scenarios admitting parametric scale separation starts with low-dimensional EFT constructions, the oldest and most famous of which being the DGKT setup \cite{DeWolfe:2005uu}. Before we start reviewing this, since most of the constructions are developed in the framework of type IIA string theory compactifications with fluxes, we first give a quick summary of the EFT machinery in this context. We assume the reader is familiar with flux compactifications in warped Calabi--Yau orientifolds and their relevance to the moduli stabilisation program in string theory. See \cite{Grana:2005jc,Douglas:2006es,Blumenhagen:2006ci,Becker:2006dvp,Marchesano:2007de,Denef:2008wq,Denef:2007pq,Ibanez:2012zz,Quevedo:2014xia,Baumann:2014nda,VanRiet:2023pnx} for introductions and reviews about the subject.

Here we will follow very similar notations as the ones used in these recent works \cite{Marchesano:2019hfb,Marchesano:2020uqz,Carrasco:2023hta} even though some conventions may vary. Focusing on four-dimensional effective theories, we consider type IIA string theory compactified on $X_4 \times X_6$ where the six-dimensional compact manifold $X_6$ admits an SU(3) structure. This structure is characterised by the usual invariant two- and three-forms $J$ and $\Omega$ from which we define the volume form to be $-\frac{1}{6}J^3 = \frac{1}{4} \im \Omega \wedge \re \Omega$. The compactification is further projected by an orientifold quotient $\Omega_p (-1)^{F_{\rm L}} {\cal R}$ \cite{Blumenhagen:2005mu,Blumenhagen:2006ci,Marchesano:2007de,Ibanez:2012zz,Marchesano:2022qbx} involving the parity-reversal worldsheet operator $\Omega_{\rm p}$, the left spacetime fermion number $F_{\rm L}$ and some involution ${\cal R}$ acting on the structure forms like ${\cal R}(J,\Omega) = -(J, \bar{\Omega})$. The question of whether the induced O6-planes wrapped around three-cycles of the internal manifold $X_6$, which are localised sources, induce a fatal departure from the SU(3) structure is actually one of the many concerns related to the AdS vacua constructions and in particular for those exhibiting an apparent separation of scales. We will come back to this issue in sect.~\ref{sec:smearing?} and proceed with the SU(3) structure from now on.

The expansion of the structure forms $(J,\Omega)$ characterises the metric deformations of $X_6$. For the two-form $J$ we build the complexified Kähler form $J_{\rm c}$ and write \cite{Gurrieri:2002wz,DAuria:2004kwe,Tomasiello:2005bp,Grana:2005ny,Kashani-Poor:2006ofe}
\begin{equation}
J_{\rm c} \equiv B+iJ = \left( b^a + i t^a\right) \om_a \, , 
\end{equation} 
from which we define the complexified K\"ahler fields $T^a \equiv b^a + i t^a$. The set of two-forms $\{\om_a\}$ denotes a basis of the $\cR$-odd cohomology group $H^{1,1}_-(X_6,\Z)$ which for simplicity we assume to span the whole cohomology group $H^{1,1}(X_6,\Z)$. The description of the most general case when $\cR$-even forms are present can be found in \cite{Marchesano:2020uqz}.

The complex structure deformations are defined through $\Omega$ with the expansion
\be
\Omega \equiv {\cal Z}^\kappa \alpha_\kappa - {\cal F}_\lambda \beta^\lambda\, ,
\ee
where now $(\alpha_\kappa, \beta^\lambda)$ is a symplectic basis of three-forms which we split into ${\cal R}$-even $(\alpha_K, \beta^\Lambda)$ and ${\cal R}$-odd  $(\beta^K, \alpha_\Lambda)$ three-forms\footnote{In the case of a Calabi--Yau geometry the set of $p$-forms $\{ \ell_s^{-2} \omega_a,  \ell_s^{-3} \alpha_\kappa,  \ell_s^{-3} \beta^\lambda, \ell_s^{-4} \tilde{\omega}^b\}$ introduced so far corresponds to a basis of harmonic forms that are integrally quantised. Moreover they form a symplectic basis, i.e. they are chosen such that $\int \omega_a \wedge \tilde{\omega}^b=\ell_s^6\delta_a^b$ and  $\int \alpha_\kappa \wedge \beta^\lambda=\ell_s^6 \delta_\kappa^\lambda$. In the more general case of an SU(3)-structure geometry some of these forms lose their harmonic property (see more about compactifications on SU(3)-structure spaces in \cite{Grana:2005ny,Kashani-Poor:2006ofe}). The definition of quantised $p$-form is then more intricate but can still be made precise in terms of smeared delta forms \cite{Casas:2023wlo}.}. A compensator $\mathcal{C}\equiv e^{-\phi}e^{\frac{1}{2}(K_{\rm cs}-K_K)}$ \cite{Grimm:2004ua} is then used to define the complexified 3-form $\Omega_c$ as
\begin{equation}
    \Omega_c\equiv C_3+i \im (\mathcal{C}\Omega)\, ,
\end{equation}
 where $\phi$ is the 10$d$ dilaton, $K_{\rm cs} \equiv - \log \left(i\ell_s^{-6} \int_{X_6} \ov \Omega \wedge   \Omega \right)$ and $K_K$ is defined as in \eqref{KK}. The complex-structure deformations then read
\begin{equation}\label{cpxmoduli}
N^K \equiv \xi^K + i n^K = \ell_s^{-6} \int_{X_6} \Omega_c \wedge \beta^K\, , \qquad  U_{\Lambda} \equiv \xi_\Lambda + i u_\Lambda =  \ell_s^{-6} \int_{X_6} \Omega_c \wedge \alpha_\Lambda\, .
\end{equation}
It is useful to gather the moduli $\{N^K, U_\Lambda\}$ into a single $U^\mu$, by defining a unified basis $\{\alpha_\mu\}\equiv \{\alpha_K,-\beta^\Lambda\}$ and $\{\beta^\mu\}\equiv \{\beta^K,\alpha_\Lambda\}$. 

The Kähler fields together with the complex-structure moduli are described by a K\"ahler potential of the form $K \equiv K_K + K_Q$, where the two pieces read
\be
K_K \,  \equiv \, -{\rm log} \left(\frac{i}{6} \CK_{abc} (T^a - \bar{T}^a)(T^b - \bar{T}^b)(T^c - \bar{T}^c) \right) \, = \,  -{\rm log} \left(\frac{4}{3} \cK\right) \, ,
\label{KK}
\ee
and 
\begin{equation}\label{KQ}
 K_Q \equiv -2 \log \left( \frac{1}{4} \RE({\cal C}{\cal Z}^\Lambda) \IM({\cal C} {\cal F}_\Lambda) - \frac{1}{4} \IM({\cal C} {\cal Z}^K) \RE({\cal C} {\cal F}_K) \right) = 4\phi_4 \, .
\end{equation}
In the formulas, we have defined ${\cal K}_{abc} \equiv - \ell_s^{-6} \int_{X_6} \omega_a \wedge \omega_b \wedge \omega_c$ which denote the triple intersection numbers of $X_6$, $\cK \equiv \cK_{abc} t^at^bt^c \equiv 6 {\rm Vol}_{X_6}$ and eventually $\phi_4 \equiv \phi - \oh \log {\rm Vol}_{X_6}$ is the 4$d$ dilaton. 

From there, restricting to the case where the smeared metric (see sect.~\ref{sec:smeared}) for $X_6$ is Calabi--Yau, one can introduce RR and NSNS three-form fluxes to generate a Landscape of vacua and exhibit infinite families of AdS$_4$ solutions with interesting properties. We use the democratic formulation of type IIA supergravity \cite{Bergshoeff:2001pv}, in which all RR potentials are grouped in a polyform ${\bf C} = C_1 + C_3 + C_5 + C_7 + C_9$, as well as their gauge invariant field strengths defined as
\be
{\bf G} \,\equiv \, \dd_H{\bf C} + e^{B} \wedge {\bf \bar{G}} \, ,
\label{bfG}
\ee
with $H=\dd B$ the three-form NSNS flux, $\dd_H \equiv (\dd - H \wedge)$ is the $H$-twisted differential and ${\bf \bar{G}}$ a formal sum of closed $p$-forms on $X_6$. It is also possible to supplement this construction with additional geometric and non-geometric fluxes with the introduction of a more generic twisted differential operator $\mathcal{D}$ \cite{Shelton:2006fd} from which the superpotential is built:
\begin{equation}
\mathcal{D}\equiv d-H\wedge+\, f\triangleleft+\, Q\triangleright+\, R\bullet\, .
\end{equation}
Here $f$ describes geometric fluxes, or metric fluxes, while $Q$ and $R$ encode the non-geometric fluxes, global and local respectively \cite{Wecht:2007wu,Plauschinn:2018wbo}. The strange symbols simply denote the action of the various fluxes when the generalised differential operator acts on the basis of forms. The usual $H\wedge$ of course takes a $p$-form to a $(p+3)$-form, the metric-flux action $f\, \triangleleft$ takes a $p$-form to a $(p+1)$-form, $Q\, \triangleright$ takes a $p$-form to a $(p-1)$-form and  finally $R\, \bullet$ transforms a $p$-form into a $(p-3)$-form. The precise definition of these actions can be found in \cite{Marchesano:2020uqz}. 

The fluxes then enter the perturbative superpotential $W\equiv W_{\rm RR} + W_{\rm NS}$, where $W_{\rm RR}$ involves K\"ahler fields and RR fluxes \cite{Taylor:1999ii}
\begin{equation} \label{WRR}
\ell_s W_{\rm RR} = - \frac{1}{\ell_s^5} \int_{X_6} {\ov{\bf G}} \wedge e^{J_c}= e_0 + e_a T^a + \frac{1}{2} {\cal K}_{abc} m^a T^b T^c + \frac{m}{6} {\cal K}_{abc} T^a T^b T^c\,,
\end{equation}
while $W_{\rm NS}$ contains the complex structure fields and the $H$-flux quanta as well as the geometric and non-geometric fluxes
\begin{equation} \label{WNS}
\ell_s W_{\rm NS} = -\frac{1}{\ell_s^5} \int_{X_6}\Omega_c\wedge \mathcal{D}\left(e^{J_{\rm c}}\right) = U^\mu\left[h_\mu+f_{a\mu}T^a+\half\K_{abc}T^bT^c\tensor{Q}{^a_\mu}+\frac{1}{6}\K_{abc}T^aT^bT^cR_\mu\right]\, ,
\end{equation}
where we have defined for compactness $h_K N^K  + h^\Lambda U_{\Lambda} \equiv h_\mu U^\mu$. The quantities $e_0, e_a, m^a, m, h_\mu$ and  $f_{a\mu}, \tensor{Q}{^a_\mu}, R_\mu$ are all integers, defined following the same conventions as in \cite{Marchesano:2021ycx}
\begin{equation}
\begin{gathered}
m \, = \,  \ell_s G_0\, ,  \qquad  m^a\, =\, \frac{1}{\ell_s^5} \int_{X_6} \ov{G}_2 \wedge \tilde \omega^a\, , \qquad  e_a\, =\, - \frac{1}{\ell_s^5} \int_{X_6} \ov{G}_4 \wedge \omega_a \, ,\\
\quad e_0 \, =\, - \frac{1}{\ell_s^5} \int_{X_6} \ov{G}_6 \,,\qquad 
h_\mu=\frac{1}{\ell_s^5}\int_{X_6} H\wedge \alpha_\mu\, ,
\end{gathered}
\label{RRfluxes}
\end{equation}
while the definition of the geometric and non-geometric fluxes can be found in \cite{Marchesano:2020uqz}.

Under the assumption that these background fluxes do not induce large corrections to the Kähler potential, the F-term scalar potential $V_F$ reads
\begin{equation}
\kappa_4^2V_F=e^K\left(K^{\A\ov \B}D_\A W\ov D_{\ov\B}\ov W-3|W|^2\right)\, ,
\end{equation}
where both the Kähler and complex-structure moduli are gathered in the index $\A\equiv\{a,\mu\}$ and where $D_\A W\equiv\partial_\A W+(\partial_\A K)W$. Note that no D-term arises thanks to our simplifying assumption that the $\mathcal{R}$-odd cohomology group $H^{1,1}_-(X_6,\Z)$ spans the whole cohomology group $H^{1,1}(X_6,\Z)$. In general, when $H^{1,1}_+(X_6,\Z)$ is not trivial, a D-term is generated due to the presence of geometric and non-geometric fluxes \cite{Ihl:2007ah,Robbins:2007yv} (see \cite{Marchesano:2020uqz} for a vacua classification analysis including such a D-term).

\subsection{Scale-separated scenarios}

We will now go through the constructions developed in the literature in the effective type IIA framework sketched above to achieve a parametric separation of scales.

\subsubsection{Vanilla DGKT construction}
\label{sec:vanilla}

The first proposition for building AdS$_4$ vacua with parametric scale separation was put forward in \cite{DeWolfe:2005uu}, relying on earlier type IIA analyses \cite{Villadoro:2005cu,Derendinger:2005ph,Behrndt:2004km,Behrndt:2004mj}. The setup established here is now commonly dubbed DGTK after the names of the authors. The framework of the analysis is the one described in the previous subsection with $X_6$ a generic Calabi--Yau manifold with all fluxes \eqref{RRfluxes} turned one and in particular a non-zero Romans mass $m$. The Kähler and complex-structure moduli are fully fixed in this generic setup and the analysis of the separation of scales performed in a specific toroidal $T^6/(\Z_3\times\Z_3)$ compactification that we will review now generalises to arbitrary Calabi--Yau's.

With our notations\footnote{In particular a conventional minus sign differs in the definition of the flux quanta $e_a$.}, the DGKT solution for the untwisted saxions\footnote{We will not discuss the twisted moduli for brevity.} of the $T^6/(\Z_3\times\Z_3)$ model reads
\begin{equation}
\label{eq:DGKT}
t^a\propto\frac{\sqrt{|\hat e_1\hat e_2\hat e_3|}}{|\hat e_a|\sqrt{m}}\quad\text{ with }\quad \hat e_a\equiv e_a-\half\frac{\K_{abc}m^bm^c}{m}\, .
\end{equation}
The hatted fluxes are defined to be invariant under shifts of the axions and the latin indices can take the values $1$, $2$ and $3$. From this expression, one quickly realises that large volumes are achieved whenever the fluxes $\hat e_a$ are much bigger than the Romans mass. Crucially, the flux quanta $e_a$ and $m^a$ are not constrained by any tadpole equation such that they can consistently be scaled up. Indeed the only non-trivial Bianchi identity concerns $\dd G_2$ and is obtained from the $C_7$ equation of motion. It is sourced by the kinetic term $|\tilde G_2|^2$ in the massive supergravity action \cite{Romans:1985tz}, with $\tilde G_2\equiv G_2+G_0B$, as well as by the action $S_{\rm O6}$ of the O6-planes with tensions $T_{\rm O6}$ and charges $Q_{\rm O6}$ which wrap 3-cycles:
\begin{equation}
S_{\rm O6}=-T_{\rm O6}\int \dd^7\xi e^{-\phi}\sqrt{-g}+Q_{\rm O6}\int C_7\, .
\end{equation}
The Bianchi identity then reads 
\begin{equation}
\dd G_2=G_0H+Q_{\rm O6}\delta_{{\rm O}6}\, ,
\end{equation}
where $\delta_{{\rm O}6}$ is a three-form localised on the O6-planes. When integrated it gives $mh_3=-2$ in the toroidal DGKT setup, where $h_3$ is the flux quantum associated with $H_3$. The Romans mass $m$ and NSNS flux $h_3$ are thus the only quantities that are constrained by the tadpole and are used to saturate the bound such that no D6-brane is introduced. For the other fluxes, the simplest scalings\footnote{We will review richer scalings in sect.~\ref{sec:Tdual}} considered in \cite{DeWolfe:2005uu} amount to choose homogeneous four-form fluxes $\hat e_a\sim n$. One then obtains the following behaviour for the saxions and various other quantities expressed from them:
\begin{equation}
\label{eq:DGKT2}
t^a\sim n^\half\, ,\quad \text{Vol}_{X_6}\sim n^\frac{3}{2}\, ,\quad e^\phi\propto (\K_{abc}t^at^bt^c)^{-\frac{1}{2}}\sim n^{-\frac{3}{4}}\, ,\quad e^{\phi_4}\sim n^{-\frac{3}{2}}\, .
\end{equation}
Large volumes and small string coupling are thus achieved when $n$ grows. From there we can express the precise KK and winding mass scales (without the volume as a proxy like in \eqref{eq:KKW}) which for a generic Calabi--Yau threefold take the form
\begin{equation}
\label{eq:KKW_CY}
\frac{\MK}{\Mp}\propto\frac{e^{\phi_d}}{\max \sqrt{L_a}}\, ,\qquad\qquad \frac{\MW}{\Mp}\propto e^{\phi_d}\min \sqrt{L_a}\, .
\end{equation}
where the $L_a$ refer to the sizes of the $h^{1,1}$ two-cycles of the Calabi--Yau expressed in string units. These sizes are given by the saxions $t^a$ such that in the case at hand we obtain
\begin{equation}
\label{eq:scaling_MKK}
\frac{\MK}{\Mp}\sim n^{-\frac{7}{4}}\, ,\qquad \frac{\MW}{\Mp}\sim n^{-\frac{5}{4}}\, .
\end{equation}

Using $W\propto \K_{abc}t^at^bt^c$ in the DGKT setup, the cosmological constant is evaluated to scale like
\begin{equation}
\label{eq:scaling_Lambda}
\Lambda\propto-3e^K|W|^2\sim n^{-\frac{9}{2}}\, ,    
\end{equation}
from which one derives
\begin{equation}
\label{eq:sep_DGKT}
\Rads\MK\sim n^\half\, ,\quad\text{ and }\quad \Rads\MW\sim n\, .
\end{equation}
Separation of scales is thus parametrically achieved as $n$ increases in addition to large volumes and weak coupling.

A perturbative analysis \cite{DeWolfe:2005uu} then shows that a tachyon is present in the spectrum but with a mass $m_{\rm tachyon}^2$ above the Breitenlohner--Freedman bound \cite{Breitenlohner:1982bm,Breitenlohner:1982jf} $m_{\rm BF}^2\equiv -\frac{3}{4}\kappa_4^2|V|$:
\begin{equation}
m_{\rm tachyon}^2=-\frac{8}{9}|m_{\rm BF}|^2\, ,
\end{equation}
such that no instability is generated at the perturbative level. Besides, this tachyon disappears in the supersymmetric solutions found from the 4$d$ point of view, in agreement with the generic analysis of the mass spectrum described below.

\subsubsection{Extensions and classification}

As summarised on the figure \ref{fig:map} by looking at the solid arrows originating from the DGKT node, the setup has been generalised/extended in various different ways.

A first extension (that we dub CFI for Camara, Font and Ibañez) explored at the same epoch restricts the geometry to a toroidal setting but at the same time allows for the presence of non-trivial metric fluxes  \cite{Camara:2005dc,Derendinger:2004jn,Villadoro:2005cu}. There, in particular, massive type IIA AdS$_4$ solutions with full moduli stabilisation are also uncovered, in which unconstrained flux quanta can be tuned to go to a large volume, small string coupling and scale-separated limit similarly to the DGKT setup. A motivation for the introduction of geometric fluxes is the possibility to get flux contributions to the RR tadpole of arbitrary sign, by contrast with the case without metric fluxes where this contribution has necessarily the same sign as the one coming from potential D6-branes.

These constructions also feature non-supersymmetric AdS solutions that are not delved into in the DGKT scale-separation analysis and thus constitute also an extension in that respect. In the special case of the $T^6/(\Z_3\times\Z_3)$ model, non-supersymmetric solutions have also been uncovered in \cite{Narayan:2010em}, though in slight tension with a more general Calabi--Yau based set of non-supersymmetric AdS solutions found in \cite{Marchesano:2019hfb}. This latter paper exploited a very nice factorisation property of the scalar F-term potential in type IIA between axions and saxions through a bilinear rewriting \cite{Bielleman:2015ina,Carta:2016ynn,Herraez:2018vae} that can be extended to incorporate geometric and non-geometric fluxes \cite{Marchesano:2020uqz}. This factorisation of the (F-term) scalar potential $V_F$ is as follows:
\begin{equation}
\label{eq:Vbi}
V_F\propto \rho_\A Z^{\A\B}\rho_\B\, ,
\end{equation}
where the $\rho_\A$'s are polynomials depending only on the axions and the flux quanta while the matrix $Z^{\A\B}$ only depends on the saxions. In \cite{Marchesano:2019hfb}, after a survey of possible branches of solutions thanks to this factorised expression, perturbative stability is analysed through diagonalisation of the Hessian matrix and shows universal behaviours of the mass spectrum (independent on the precise geometry), for the various non-supersymmetric and supersymmetric branches, one of the latter containing the DGKT analysis which is consistently reproduced. This is nicely summarised in \cite[Table 2]{Marchesano:2019hfb}. To be more precise here, the branch containing DGKT is dubbed ``A1-S1'' and features $h^{2,1}(X_6,\Z)$ tachyons with $m^2_{\rm tachyon}=-(8/9)|m_{\rm BF}|^2$ as well as $h^{2,1}(X_6,\Z)$ massless modes. On the other hand the non-supersymmetric version of this branch has one more tachyon. This is in perfect agreement with \cite{DeWolfe:2005uu} where the toroidal $T^6/(\Z_3\times\Z_3)$ model has no complex-structure moduli. There is then no tachyon and no massless mode in their supersymmetric vacua while a sign choice interpreted as giving rise to non-supersymmetric solutions produces one tachyon. To be complete about these solutions that generalise DGKT, the universal mass spectrum, both for the supersymmetric branch and its non-supersymmetric counterpart, reads \cite{Marchesano:2019hfb}
\begin{equation}
\label{universal_spectrum}
\frac{m_{\rm s}^2}{|m_{\rm BF}|^2}=\left(8,\frac{280}{9},8,-\frac{8}{9}\right)\, ,\quad\begin{cases}
\text{SUSY: }  &\frac{m_{\rm a}^2}{|m_{\rm BF}|^2}=(\frac{40}{9},\frac{352}{9},\frac{40}{9},0)\\
\text{non-SUSY: }  &\frac{m_{\rm a}^2}{|m_{\rm BF}|^2}=(-\frac{8}{9},\frac{160}{9},\frac{160}{9},0)\
\end{cases}\, .
\end{equation}
Here $m_{\rm s}$ and $m_{\rm a}$ stand for the masses of the axions and saxions respectively and the multiplicities of the masses are $(1,1,h^{1,1}_-(X_6),h^{2,1}(X_6))$. Note that the massless modes correspond to $h^{2,1}(X_6)$ complex-structure axions that do not appear at all in the scalar potential \eqref{eq:Vbi} and are thus true flat directions of the classical potential. Quantum corrections can then be concerning but it is possible for these massless axions to be eaten by gauge bosons living on D6-branes via a Stückelberg mechanism as described in \cite{Camara:2005dc}. A more precise case-dependent analysis would thus be required to assess the stability of these moduli at the quantum level. Introducing metric fluxes is another way to address this issue and to rely less on the aforementioned localised effects. Indeed it is shown in \cite{Marchesano:2020uqz} that the rank of the metric flux $f_{a\mu}$ determines how many complex-structure axions are stabilised.

The bilinear factorisation of the scalar potential has also been extensively used in \cite{Marchesano:2020uqz}, including all kinds of geometric and non-geometric fluxes, as an attempt to systematically classify all type IIA AdS solutions in a common framework. As it happens, this classification missed some important massless type IIA branches that can be understood from T-duality (see sect.~\ref{sec:Tdual}) and which exhibit scale separation \cite{Cribiori:2021djm}, but the discussion has been completed in \cite{Carrasco:2023hta}.

\subsubsection{Lower dimension and type IIB solutions}
\label{sec:3d}

AdS scale separation has also recently been extensively explored in lower dimensions and in particular in 3$d$ thanks to compactifications on manifolds with $G_2$ holonomy \cite{Farakos:2020phe,VanHemelryck:2022ynr,Farakos:2023nms,Farakos:2023wps}. A nice motivation for studying these less phenomenologically appealing AdS$_3$ solutions is that they are dual to two-dimensional CFT's from an holographic viewpoint, which are simpler than higher-dimensional ones, and could thus provide insights on how scale separation is realised in the dual frame (see sect.~\ref{sec:holo} for more on the relation between scale separation and holography). Multiplying diverse constructions is also a way to probe more widely the various swampland conjectures that have consequences on scale separation (see sect.~\ref{sec:conjs}).

The question of parametric scale separation in type IIB string theory has also been investigated, in 3$d$ again in \cite{Emelin:2021gzx}, where some hints of scale separation have been put forward although proper quantisation of the fluxes there are an obstacle to reach conclusive answers. In 4$d$, type IIB AdS solutions based on SU(2)-structure spaces and with scale separation have been described in \cite{Caviezel:2009tu,Petrini:2013ika} but the more recent analysis \cite{Cribiori:2021djm} (though focused on type IIA vacua) identified vanishing cycles in the scale-separated regime, that consequently seem to spoil the aforementioned constructions. As of today then, no conclusive scale-separated constructions in type IIB have been built (recall that we exclude KKLT vacua from our discussion, see sect~\ref{sec:casimir} for a very brief comment).

\subsubsection{T-dual frame and massless type IIA solutions}
\label{sec:Tdual}

For now we still stay in the realm of low-dimensional constructions and we will explore another set of proposed scale-separated solutions that has recently been updated and upgraded. These constructions are depicted on the map \ref{fig:map} in the ``T-dual corner'' as they can be seen to be related to the ones mentioned so far via T-duality. Before describing the various works that have been done along these lines, we review the generic discussion presented in \cite{Carrasco:2023hta} to understand how more generic flux scalings than the ones considered in the DGKT setup can still allow for scale separation while naturally giving rise to a T-dual picture. We also explain the consequences of the T-duality on the fluxes to understand the ingredients present in these kind of solutions.

We start again from the DGKT solutions \eqref{eq:DGKT}, valid for compactifications on toroidal orbifolds of the form $(T^2)^3/\Gamma$ where $\Gamma$ denotes an orbifold group action. Remember that the DGKT solutions involved an homogeneous scaling of the fluxes $\hat e_a\sim n$. We can generalise this in two ways: By scaling asymmetrically the fluxes\footnote{Anisotropic scalings have also been explored in \cite{Tringas:2023vzn}.} and by allowing arbitrary powers. We single out the flux quantum $\hat e_1$ and choose
\begin{equation}
\hat e_1\sim n^{2r}\, ,\qquad\hat e_2,\hat e_3\sim n^{r+s}\, ,
\end{equation}
where $r$ and $s$ are free parameters. The DGKT scalings then simply correspond to the choice $r=s=\half$. With these asymmetric scalings, the sizes of the two-cycles scale differently, like
\begin{equation}
\label{eq:scalings_gen}
t^1\sim n^s\, \qquad t^2,t^3\sim n^r\, .    
\end{equation}
The overall volume and the 10$d$ and 4$d$ dilatons \eqref{eq:DGKT2} now scale like
\begin{equation}
{\rm Vol}_{X_6}\sim n^{2r+s}\, ,\quad e^\phi\sim n^{-r-\frac{s}{2}}\, ,\quad e^{\phi_4}\sim n^{-2r-s}\, ,
\end{equation}
from which we deduce the following behaviour for the KK and winding scales:
\begin{equation}
\frac{\MK}{\Mp}\sim n^{\min\{-2r-\frac{3s}{2},-\frac{5r}{2}-s\}}\, \qquad \frac{\MW}{\Mp}\sim n^{\min\{-2r-\frac{s}{2},-\frac{3r}{2}-s\}}\ .
\end{equation}
The cosmological constant reads
\begin{equation}
\Lambda\sim n^{-6r-3s}\ ,    
\end{equation}
so that we finally obtain
\begin{equation}
\Rads\MK\sim n^{\min\{r,\half(r+s)\}}\, ,\quad\Rads\MW\sim n^{\min\{r+s,\half(3r+2s)\}}\, .
\end{equation}

With these behaviours, scale separation is achieved whenever we have $r>0$ and $r+s>0$. This also ensures a large volume regime and in addition, going to weak coupling requires to have $r+\frac{s}{2}>0$. Crucially, these conditions can still be satisfied even when $s$ is negative\footnote{In this case we see from \eqref{eq:DGKT} that to preserve an exact scaling symmetry and satisfy flux quantisation the flux quanta $m^1$ needs to be put to zero since otherwise it would inconsistently grow like $n^s$.}, which corresponds to a shrinking of the torus factor associated to $t^1\sim n^s$. Such scalings are then best understood after a double T-duality on the torus factor, that flips the sign of $s$ and brings us back to large volume asymptotics.

We can then get an idea of where the double T-duality brings us by looking at how it affects the effective 4$d$ theory \cite{Banks:2006hg,Carrasco:2023hta}. The Kähler potential $K_K$ takes the form
\begin{equation}
K_K=-\log\left(i\kappa(T^1-\bar T^1)(T^2-\bar T^2)(T^3-\bar T^3)\right)\, ,   
\end{equation}
with $\kappa\equiv\K_{123}$ the only non-vanishing intersection number, while $K_Q$ does not depend on the Kähler moduli. The double T-duality implemented by $T^1\mapsto-\frac{1}{T^1}$, which does not affect the four-dimensional dilaton, then transforms the Kähler potential as $K_K\mapsto K_K+\log|T^1|^2$. To go back to the original form of the potential, one can then perform a Kähler transformation as follows:
\begin{equation}
K_K\mapsto K_K-F-\bar F\, ,\qquad W\mapsto e^F W\, ,    
\end{equation}
where $F\equiv\log T^1$.

After this transformation, the only difference lies in the new superpotential and we can now inspect it to infer the fluxes present in this T-dual frame. Using $\cK_{123}=\kappa$ as the only non-vanishing intersection number in our factorised torus orbifold, one can easily perform the T-duality in the superpotentials \eqref{WRR}, \eqref{WNS} (expressed without the metric and non-geometric fluxes) and multiply by $e^F=T^1$ to obtain
\begin{subequations}
\label{newW}
\begin{align}
\ell_s W_{\rm RR}  &= T^1\left(e_0 + e_2T^2+e_3T^3 + \kappa m^1T^2T^3\right) - e_1 -\kappa m^2T^3 - \kappa m^3 T^2 - \kappa mT^2T^3\, , 
\label{newWRR} \\
\ell_s W_{\rm NS} & =  h_\mu U^\mu T^1\, .
\label{newWNS}
\end{align}
\end{subequations}
The most striking feature of this T-dual superpotential is that there are no cubic terms anymore in $W_{\rm RR}$ and hence no Romans mass is present if we have $m^1=0$ in the original frame to keep an exact scaling symmetry of the equations. Besides, the linear structure of $W_{\rm NS}$ in $U^\mu$ is replaced by a quadratic term mixing it with $T^1$ which indicates that the $H$-flux has been replaced by a rank-one metric flux. The generic scalings \eqref{eq:scalings_gen} when $s$ is negative are thus understood through T-duality in the simplest cases as massless type IIA solutions without NSNS $H$-flux but twisted with geometric fluxes. Having no Romans mass is naively a good news for the quest of understanding the higher-dimensional uplift of the solutions (see sect.~\ref{sec:smearing?}) but the price to pay is to depart from a Calabi--Yau geometry.

The T-duality has first been performed precisely in this toroidal context in \cite{Banks:2006hg} and further investigated in \cite{Caviezel:2008ik} and more recently in \cite{Cribiori:2021djm} where the authors build scale-separated solutions in this setup. They also analyse the backreaction of the orientifold planes and investigate an M-theory uplift but that will be discussed further in sect.~\ref{sec:smearing?}. From the 4$d$ point of view, infinite families of vacua featuring scale separation and weak coupling can then be found, similarly to the DGKT solutions but here without Romans mass. In \cite{Carrasco:2023hta}, new families were built by generalising the toroidal framework above. Intuitively, what is required to implement the same kind of double T-duality is a perturbative $SL(2,\Z)$ duality on one Kähler modulus which naturally leads to consider smooth elliptically fibred Calabi--Yau's as candidates for generalising the construction. The elliptic fibre plays the role of the above $T^2$ factor associated to $t^1$ which scales like $n^s$ and leads to a small fibre area when $s$ is negative. A double T-duality on the fibre, or more precisely a Fourier--Mukai transformation in this case, is expected to similarly map the setup to a large-volume massless type IIA configuration twisted by metric fluxes. The study developed in \cite{Carrasco:2023hta} makes use of the same classification techniques of \cite{Marchesano:2020uqz} and indeed finds branches that were missed and that precisely contain these generic massless T-dual solutions. It is also shown that despite the more involved structure of the triple intersection numbers compared to the toroidal case, one can still build infinite families of scale-separated vacua. The scaling symmetry exhibited in \cite{Cribiori:2021djm} is no longer exact but still allows for scale separation asymptotically where it is recovered. The perturbative stability of these new families has been checked and the mass spectrum is shown to asymptotically match the universal DGKT one mentioned in \eqref{universal_spectrum}. Departure from the universal behaviour at low $n$ is thought to arise because of curvature corrections expected to show up on the T-dual DGKT side due to kinematic mixing of the small Kähler modulus with the others. These corrections should be incorporated in the T-duality and they explain the deviations from the universal mass spectrum. At large $n$ however, the modulus still decouples from the rest and the correct spectrum is found, as in the trivially factorised cases for which the modulus is decoupled for any $n$. The absence of non-perturbative instabilities has also been checked and at worse only marginal membranes have been found.

\subsubsection{Quantum corrections and scale separation}
\label{sec:casimir}

We focused so far on parametric scale separation obtained at the classical level but it is worth mentioning that scenarios exist to achieve scale separation thanks to quantum corrections. The KKLT \cite{Kachru:2003aw} and LVS \cite{Balasubramanian:2005zx} constructions are examples using quantum corrections to achieve a scale-separated AdS vacuum as a first step before uplifting to dS. However as we already briefly mentioned, these scenarios do not offer a proper parametric control to generate clean infinite families of scale-separated vacua, which is what this review focuses on. It is still interesting to mention that despite difficulties to achieve a small $W_0$ in KKLT, synonym of scale separation, some works have dug in that direction \cite{Demirtas:2019sip,Demirtas:2020ffz,Demirtas:2021nlu,Demirtas:2021ote,Grana:2022nyp}.

Parametric scale separation in AdS$_4$ has however been achieved by using quantum corrections in the framework of M-theory in \cite{Luca:2022inb}. The authors define an energy condition called Reduced Energy Condition (REC) \cite{DeLuca:2021mcj} which constrains the stress-energy tensor in the context of compactification and which is satisfied for a large set of matter content. Violating the REC is expected to help building scale-separated solution, in agreement with \cite{Gautason:2015tig} where scale separation is ruled out in absence of negatively-charged localised objects which violate the REC (more on this in sect.~\ref{sec:unavoidable}). In \cite{Luca:2022inb}, contrary to DGKT-like constructions which involve O-planes, the authors propose to violate the energy condition with quantum effects by considering the Casimir energy on the internal space. M-theory is compactified down to 4$d$ on an isotropic torus $T^7$ and to get a non-vanishing Casimir energy, supersymmetry is broken à la Scherk--Schwarz \cite{Scherk:1978ta,Scherk:1979zr,Rohm:1983aq,Kounnas:1988ye,Ferrara:1987es,Ferrara:1988jx,Kounnas:1989dk} with anti-periodic boundary conditions for fermions along the torus cycles. To stabilise the torus that the Casimir energy wants to expand, a seven-form flux $F_7$ in introduced, properly quantised in terms of some integer $n$ like
\begin{equation}
\frac{1}{(2\pi\ell_{11})^6}\int F_7=n\, ,
\end{equation}
where $\ell_{11}$ is the eleven-dimensional Planck length. In the large-$n$ limit the AdS and internal radii are shown to solve the 11$d$ Einstein equations at large values in Planck units such that other corrections are suppressed. From these radii the authors then deduce
\begin{equation}
\label{eq:n6}
\Rads\MK\sim n^6\, ,    
\end{equation}
and thus scale separation is parametrically achieved. However, being non-supersymmetric this construction is expected to be unstable at some level (see sect.~\ref{sec:unstab}).\\

This ends our tour of the constructions proposed in the literature for vacua with parametric scale separation built from an effective low-dimensional perspective. On the picture \ref{fig:map}, we have covered the bottom part. These bottom-up constructions are a fundamental pillar for addressing the question of scale separation in string theory but are only a modest first step. Indeed, one can only be convinced by the existence of vacua with appealing properties if they are proved to emerge from fully-fledged higher-dimensional embeddings. Works in this direction that bring pieces of evidence towards both answers (feasibility or not of such embeddings) are very numerous and our purpose is to review them now. 

\section{Higher-dimensional perspective}
\label{sec:higher}

\subsection[A 10$d$ excursion: The smearing approximation]{\bm A 10$d$ excursion: The smearing approximation}
\label{sec:smeared}

As a first approach into the higher-dimensional realm we start by defining the \emph{smearing approximation}, which makes a connection between the low-dimensional solutions and the 10$d$ equations of motions. To understand this connection, it is useful to review the features of a class of AdS$_4$ type IIA solutions compactified on SU(3)-structure geometries from a 10$d$ perspective. 

\subsubsection{Type IIA on SU(3)-structure manifolds}

AdS$_4$ solutions of type IIA supergravity on SU(3)-structure manifolds have been extensively described from a 10$d$ point of view in \cite{Behrndt:2004km,Behrndt:2004mj,Lust:2004ig,Grana:2005jc,Martucci:2005ht,Grana:2006kf,Koerber:2010bx}. The ten-dimensional background is a warped product of spacetime and the six-dimensional internal space
\begin{equation}
\dd s^2=e^{2A}g_{\mu\nu}\dd x^\mu\dd x^\nu+\dd s_6^2\, ,
\end{equation}
and the solution involves in the end a constant warp factor and dilaton, while the various fluxes involved in the setup are parameterised like
\begin{equation}
\begin{gathered}
H=2m\Re\Omega\, ,\qquad\quad e^\phi G_0=5m\, ,\qquad\quad e^\phi G_2=\frac{\tilde m}{3}J+i\W_2\, ,\\
e^\phi G_4=\frac{3m}{2}J\wedge J\, ,\qquad\quad e^\phi G_6=-\frac{\tilde m}{2}J\wedge J\wedge J\, ,
\end{gathered}
\end{equation}
where $\W_2$ is the usual second torsion class and $\tilde m$ is linked to the first class $\W_1$. Note that in these expressions, we have absorbed the constant warp-factor-dependent quantity $e^{-A}=e^{-\phi/3}$ in the Romans mass $m$ and in $\tilde m$. Besides, the two-form and three-form $(J,\Omega)$ satisfy
\begin{equation}
\dd J=2\tilde m\Re\Omega\, ,\qquad\quad\dd\Omega=-\frac{4}{3}i\tilde mJ\wedge J+\W_2\wedge J\, .
\end{equation}

As already seen in sect.~\ref{sec:vanilla}, the Bianchi identity for the $G_2$ flux coming from the $C_7$ equations of motion is $\dd G_2=G_0H+Q_{\rm O6}\delta_{\rm O6}$ where the localised three-form $\delta_{\rm O6}$ denotes Poincaré duals of O6-planes. Inserting the expression for $G_2$ above one can deduce
\begin{equation}
\label{eq:bianchi}
\dd\W_2=\frac{2i}{3}\left(\tilde m^2-15m^2\right)\Re\Omega-ie^\phi Q_{\rm O6}\delta_{\rm O6}\, . 
\end{equation}
From this relation together with the expression for $\dd\Omega$, using the fact that $\W_2\wedge\Omega=0=\dd(\W_2\wedge\Omega)$ and other useful relations \cite{Lust:2004ig}, one can derive the following identity \cite{Acharya:2006ne}:
\begin{equation}
|\W_2|^2=\frac{2}{3}\left(\tilde m^2-15m^2\right)+e^\phi Q_{\rm O6}{\rm Vol}\Sigma\, , 
\end{equation}
where $\Sigma$ formally denotes the 3-cycles wrapped by O6-planes. In \cite{Acharya:2006ne}, this relation has been used to argue that the particular case of Calabi--Yau geometries is not consistent with a non-vanishing Romans mass. Indeed, positive definiteness of $|\W_2|^2$ implies that $\tilde m^2\geq 15m^2$ such that in the Calabi--Yau limit where $\tilde m=0$, one necessarily obtains a vanishing mass parameter $m$. In particular, it shows that the DGKT-like solutions cannot be trivially uplifted to Calabi--Yau geometries from the 10$d$ point of view and one must find another way to make sense of the 4$d$ analysis that relies on a Calabi--Yau background. Actually, as we will see later, not even a generic SU(3)-structure geometry with localised sources can accommodate an uplift of DGKT. 

\subsubsection{Smeared solutions}

\begin{figure}[!ht]
    \centering
    \includegraphics[scale=0.8]{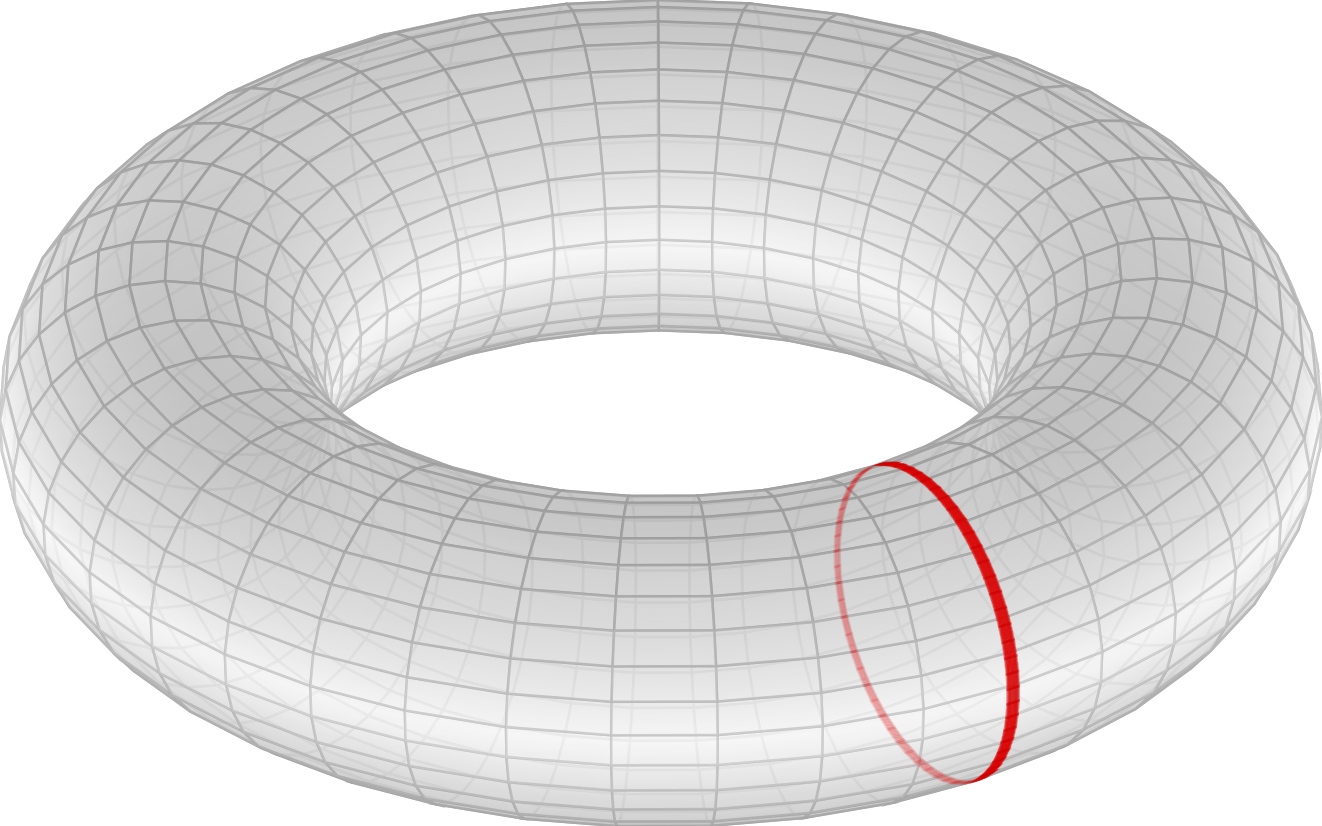}
    \quad
    \raisebox{1.7cm}{$\text{smearing}\atop \xrightarrow{\hspace{1cm}}$}
    \quad
    \includegraphics[scale=0.8]{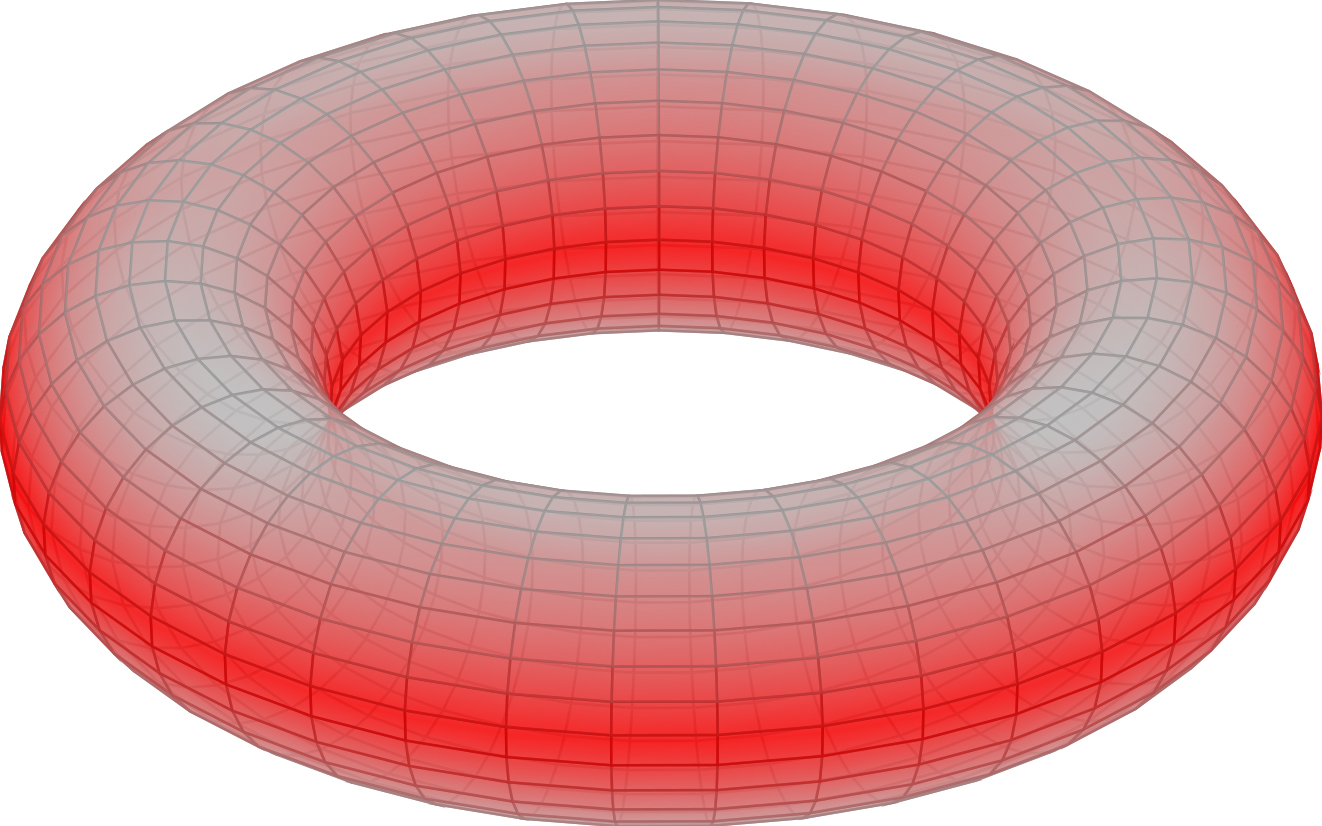}
    \caption{Schematic representation of what smearing means. In this 2$d$ representation, the localised O-plane on the left in red which threads a 1-cycle is diluted over the whole compact space on the right.}
    \label{fig:smearing}
\end{figure}

A solution to the problem mentioned above has been provided in the same work \cite{Acharya:2006ne} to justify how the Calabi--Yau starting point in the low-dimensional construction can make sense. The idea is to think the O6-planes as being \emph{smeared}, i.e. diluted, in the compact dimensions instead of being localised. The figure \ref{fig:smearing} illustrates this procedure.

The localised source form $\delta_{\rm O6}$ is thus replaced by the harmonic representative of its cohomology class, constant in the compact space:
\begin{equation}
e^{\phi}Q_{\rm O6}\delta_{\rm O6}\longrightarrow -10 m^2\Re\Omega\, .
\end{equation}
With this form, the Bianchi identity \eqref{eq:bianchi} is satisfied when $\tilde m=\W_2=0$ and now corresponds to the Calabi--Yau geometry. The interpretation is that this should provide an accurate description at long wavelengths above the compactification scale and thus reproduce the effective low-dimensional description. An important remark at the heart of some criticisms raised against the smearing approximation (see sect.~\ref{sec:smearing?}) is to realise that the topology of the 3-cycles wrapped by the O6-planes changes when one goes from the localised picture to the smeared one. Indeed, it can be shown \cite{Acharya:2006ne} that localised O6-planes wrap homologically trivial three-cycles, as a consequence of the Bianchi identity and the exactness of $\Re\Omega$, which implies that the localised charges must cancel among each other. On the other hand it is not the case in the the smeared approximation. 

With this definition of how a smeared solution is interpreted in the massive DGKT setup, we went upwards on fig.~\ref{fig:map}. Continuing at this level, the other constructions proposed for scale separation can also be understood in this smeared framework. The massless toroidal T-dual setup of \cite{Banks:2006hg} has been embedded into a smeared description in \cite{Caviezel:2008ik} while the more general T-dual solutions of \cite{Carrasco:2023hta} has been understood there to correspond to half-flat geometries, both falling into the class of type IIA solutions \cite{Behrndt:2004km,Behrndt:2004mj,Lust:2004ig} described above as expected. The reference \cite{Caviezel:2008ik} also provides a nice easy-to-grasp interpretation of smearing: One can imagine expanding the localised delta forms describing the O6-planes into Fourier modes (which would give equal weight to all plane waves) and truncating to the zero (constant) mode only, in agreement with the KK reduction philosophy to describe an EFT. Informally, if we denote by $\vec y$ the coordinates in the compact space we have
\begin{equation}
\delta_{\rm O6}(\vec y)\propto\sum_{\vec k}e^{i\vec k\cdot\vec y}=1+\sum_{\vec k\neq \vec 0}e^{i\vec k\cdot\vec y}\,\longrightarrow\, 1\, .
\end{equation}

From these considerations, it is clear that the smearing approximation undoubtedly provides a connection between the type IIA 10$d$ realm and the low-dimensional constructions, by showing that the EFT description only solves the zero mode of the 10$d$ equations of motions. Of course, that the connection is clear does not mean it is consistent in the sense that the smeared solutions indeed define a good approximations of fully-fledged uplifted vacua. Criticisms have been raised along these lines as well as ``counter-criticisms'' and we will review them now.

\subsection{Is painting smeared O-planes art or deception?}
\label{sec:smearing?}

As stated above, the peculiar properties of the smearing approximation to map the 10$d$ equations of motions to the 4$d$ effective description can raise doubts about its consistency to provide a solid ground upon which constructing phenomenologically interesting (in particular scale-separated) solutions. Starting from the smearing approximation, the question is thus to determine if a proper localisation of the O6-planes which induces a backreaction would kill scale separation or even simply fatally drive the smeared solutions away from any consistent effective description. Lots of works have been dedicated to these questions.

\subsubsection{An M-theory lift of DGKT attempt}
\label{sec:Mlift}

Worries have first been expressed soon after the DGKT proposal in \cite{Banks:2006hg} which is the work where the authors described the T-duality procedure that we mentioned earlier to reach massless type IIA solutions on twisted geometries. The purpose of the authors was to use the T-duality in order to then apply techniques \cite{Haack:2001jz,Singh:2001gt,Singh:2002tf} to build a M-theory uplift of DGKT. They proceed in two steps with first what they call a naive lift without taking the O-planes into account and then a refinement to introduce them in the picture. They find a consistent uplift that can be described by 11$d$ supergravity but only in a regime where only one of the three scalable fluxes grows, which is not the one where the DGKT setup has control. The controlled regime corresponds to a homogeneous scaling of the fluxes and the authors conclude that in this limit, weak coupling should not be possible everywhere on the compact manifold with an asymptotically growing region where it fails. Moreover, an attempt to resolve this strong coupling region with 11$d$ supergravity indicates that it could grow in the asymptotic limit and consequently spoil scale separation. 

We will see later how a recent work \cite{Cribiori:2021djm} came back to the issue of building an M-theory uplift in this context and managed to overcome the issues mentioned here.

\subsubsection{Arguments against smearing as a good starting point}
\label{sec:against_smearing}

Besides, a variety of arguments against the smearing approximation as being a valid first step to build vacua has been raised in \cite{McOrist:2012yc}. 
\begin{itemize}
\item First from a generic point of view, the topological change induced by the smearing approximation defined earlier is criticised. The situation contrasts with type IIB string theory where the backreaction of fluxes can sometimes preserve the Calabi--Yau structure (conformal Calabi--Yau solutions exist). To illustrate that this can be a problem, one can notice that at large volume the Romans mass does not dilute contrary to other fluxes. This makes it very unclear that it makes sense to approximate a solution with a Calabi--Yau (or even SU(3)-structure) ansatz like in the smeared approximation when the O-planes are truly localised since they cannot cancel everywhere the non-diluting energy density. In a similar spirit, it is argued in \cite{Douglas:2010rt} that the backreaction corrections must be comparable in magnitude to the fluxes in the 10$d$ equations of motion, precisely to be able to cancel the flux energy everywhere in the compact manifold.
\item  Second, the very definition of orientifold planes in a massive IIA setup is challenged as it is a very complicated task to get a proper perturbative string description in such backgrounds. The cocktail of Romans mass mixed with O-planes, which are crucial ingredients of the DGKT proposal, thus seems to rest on shaky grounds. 
\item Thirdly, the authors emphasise that no SU(3)-structure solution with constant dilaton and warp factor can satisfy the localised 10$d$ equations of motion, which was already known from \cite{Acharya:2006ne}. They notice that deformation to the most generic $SU(3)\times SU(3)$ structure case could help but again to conclude that the smeared solutions do not make sense in the DGKT setup.
\end{itemize} 

Despite these concerns, there has actually been progress towards proving that the smeared solutions indeed seem to provide a good starting point which defines the leading contribution of a true localised solution as we will see below. However before describing these attempts to expand a localised solution around the smeared solution we will review some famous no-go theorems and extensions related to scale separation that forbid (or at least put strong constraints on) ``pure-flux'' massive type IIA solutions i.e. without negative tension objects.  

\subsubsection{O-planes and their singularities seem unavoidable}
\label{sec:unavoidable}

If the orientifold planes and their localised nature lead to so many troubles one could think evading this by either removing them from the constructions or by finding ways to smooth them already at the supergravity level.

Concerning the former proposition, several no-go theorems and analyses suggest that orientifold planes and their negative tensions are unavoidable if one wants to achieve scale separation. A first famous no-go has been stated in \cite{Maldacena:2000mw} and is commonly referred to as the Maldacena--Nunez theorem. By analysing the equations of motions of massive type IIA supergravity, one finds that compactifications on smooth manifolds without boundaries cannot yield a Minkowski or de Sitter spacetime in lower dimensions. This no-go seems a priori non relevant for our discussion focused on anti de Sitter spacetimes but it has actually been slightly extended to forbid small negative cosmological constants in \cite{Gautason:2015tig}. To do so, the authors compared the magnitude of the AdS radius to that of the curvature radius $R_{\rm compact}$ of the internal space and found
\begin{equation}
\label{eq:no-go}
\Rads\lesssim R_{\rm compact}\ .
\end{equation}
Under the assumption that the KK length cannot be decoupled from the curvature radius the above inequality translates into the absence of scale separation. Note that the impossibility to decouple the KK length from the internal curvature radius is crucial for the no-go to apply or else \eqref{eq:no-go} would not be equivalent to the absence of scale separation anymore. Such absence of decoupling has been shown in \cite{Collins:2022nux} for particular Einstein geometries from an holographic point of view and has been conjectured to hold in full generality. We will see in the next subsection that potential counterexamples may exist and we will come back to the holographic statement in sect.~\ref{sec:holo}.

This result thus suggests that scale separation in AdS cannot be achieved without introducing orientifold planes and their singularities (or without large dilaton gradients). It is in agreement with \cite{Tsimpis:2012tu} where specific pure-flux (i.e. without O-planes) compactifications down to AdS$_4$ vacua in massive type IIA were shown to exhibit no scale separation despite a naive expectation that they could\footnote{Similar results have more recently been developed for AdS$_2$ compactifications \cite{Lust:2020npd}.}. Through this extended Maldacena--Nunez no-go theorem, the fact that orientifold planes are required as well as their singularities at the supergravity level seems to prevent the promising smoothing investigated in \cite{Saracco:2012wc,Saracco:2013aoa} from happening. This was already anticipated in \cite{McOrist:2012yc} and has been made explicit and precise in \cite{DeLuca:2021mcj} despite a more optimistic view exposed in \cite{Junghans:2020acz}. The smoothing procedure was a way to get rid of the O-plane singularities by resolving them already at the level of supergravity. But the absence of large dilaton gradients in this smoothing procedure and the lack of sources with negative tension seems to make the no-go theorem unavoidable in this case and the smoothing unlikely to happen.

We then learn that in our quest for scale separation, we should embrace the O-plane singularities and deal with them. If one wants to make sense of the smearing approximation, one then needs to explicitly show that it can provide a sensible approximation of localised solutions, challenging the pessimistic expectations of sect.~\ref{sec:against_smearing}.

\subsubsection{Source localisation}
\label{sec:localisation}

Along these lines, works have been devoted to explicitly evaluate the backreaction of the O6-planes on the solutions due to their localisation to check if smearing makes sense. For scale-separated vacua however, this backreaction has only been evaluated at first order in a perturbative expansion with respect to some parameter (related in the end to the unbounded flux $n$) around the smeared solutions. But going beyond the smearing approximation in this way, even at modest order in a perturbative expansion, still provides a non-trivial test of the scale-separated constructions since things could already fail at this level.

From a general point of view, not focused on scale separation, the smearing of sources (O-planes and D-branes) has been studied in \cite{Blaback:2010sj}. There it is shown that the BPSness ($\neq$ supersymmetric) of the solutions seems to drastically affect the consistency of the smearing procedure. Indeed localised solutions for BPS constructions, T-dual to GKP \cite{Giddings:2001yu}, are uncovered as well as their relations with the smeared solutions. In this case, the authors show that despite modifications induced by the localisation, the smeared approximation still captures characteristic features of the localised version. This contrasts with the same analysis performed for non-BPS solutions where localisation is impossible.

A more recent work \cite{Baines:2020dmu} analysed 7$d$ non-supersymmetric compactifications on $T^3/\Z_2$ (\cite{Blaback:2013taa}) and solved the 10$d$ equations of motions with localised O6-planes, finding no obstruction. The solution is Minkowski with flat directions such that AdS scale separation is not truly achieved and the stringy fate of the vacuum is not known, but the construction still challenges the criticisms reviewed in sect.~\ref{sec:against_smearing} by showing that weak coupling and large volume limits can be taken and reproduce the smeared solution away from the orientifold planes. The authors also oppose the lore saying that the non-dilution of the Romans mass constitutes an obstacle to define a consistent smeared Calabi--Yau limit. Indeed, they argue that the zero-form flux, even if it is not dressed with inverse metric factors, is still dressed with some $g_{\rm s}$ power such that when the large volume and weak coupling limits are taken together, the Romans mass actually dilutes exactly like the other fluxes and thus smearing makes sense. All this is not proof that localised solutions truly exist (even less for the phenomenologically interesting scale-separated solutions like DGKT) since the O-plane singularities would still need to be resolved by string effects in M- of F-theory for example. But at least as far as the massive 10$d$ supergravity equations of motion are concerned: O-plane localisation works just fine in this example. One caveat emphasised by the authors though is the absence in their setup of O-plane intersections (in the covering space) that could bring up more problems due to matter localised there and seem to deserve a dedicated treatment. We will see in sect.~\ref{sec:intersection} that this may actually not be the case.\\

Similarly two recent works tackled the question of source localisation for the specific scale-separated DGKT solutions with optimistic conclusions. Indeed in \cite{Junghans:2020acz} the author builds exact solutions to the localised 10$d$ equations of motion through a perturbative expansion around the smeared solution. The idea is precisely to test if the full solution can be expressed as the smeared one corrected by small contributions that are subleading in the large-$n$ limit. More precisely, the various fields are expanded around the smeared solution indicated with ``$(0)$'' indices below as (note that the scalings are written explicitly such that the smeared coefficient as well as the corrections do not scale)
\begin{equation}
\begin{aligned}
&G_6=0+G_6^{(1)}n+\mathcal{O}(n^{\half})\, ,\quad &&G_4=G_4^{(0)}n+G_4^{(1)}n^\half+\mathcal{O}(1)\, ,\\ &G_2=G_2^{(0)}n^\half+G_2^{(1)}+\mathcal{O}(n^{-\half})\,\quad &&H=H^{(0)}+H^{(1)}n^{-\half}+\mathcal{O}(n^{-1})\, ,\\
&g_{\rm s}=g_{\rm s}^{(0)}n^{\frac{3}{4}}+g_{\rm s}^{(1)}n^{-\frac{1}{4}}+\mathcal{O}(n^{-\frac{5}{4}})\, ,\quad &&A=A^{(0)}n^{\frac{3}{4}}+A^{(1)}n^{\frac{1}{4}}+\mathcal{O}(n^{-\frac{5}{4}})\, ,\\
&g_{mn}=g_{mn}^{(0)}n^\half+g_{mn}^{(1)}n^{-\half}+\mathcal{O}(n^{-\frac{3}{2}})\, ,
\end{aligned}
\end{equation}
where $A$ is the warp factor and $g_{mn}$ the six-dimensional metric, Ricci-flat at the smeared level.

These first-order expansions are plugged into the equations of motion and everything works fine even if corrections to the equations themselves are not subleading (in agreement with \cite{Douglas:2010rt} cited earlier). In addition, several corrections to the 4$d$ scalar potential are evaluated and shown to drop out at large $n$ to conclude that moduli stabilisation and scale separation should work as in the DGKT setup\footnote{Note that all the derivation is expected to hold similarly for the non-supersymmetric extensions of DGKT.}. These corrections are nicely summarised in \cite[Table 3]{Junghans:2020acz} that we copy here in table~\ref{tab:junghans}.

\begin{table}[!ht]
    \centering
    \begin{tabular}{|l|l|}
    \hline
    Corrections to the 4$d$ scalar potential & Order\\\hline
    Leading potential &  $n^{-\frac{9}{2}}$\\
    Backreaction in the bulk & $\lesssim n^{-5}$\\
    $\alpha'$ in the bulk (8-derivative) & $\lesssim n^{-5}$\\
    $\alpha'$, $g_{\rm s}$ backreaction in the tube regions\tablefootnote{These regions refer to tubes around the O6-planes whose size decreases at large-$n$.} & $\lesssim n^{-\frac{21}{4}}$\\
    $g_{\rm s}$ in the bulk (8-derivative, 1-loop) & $\lesssim n^{-\frac{13}{2}}$\\\hline
    \end{tabular}
    \caption{Table extracted from \cite{Junghans:2020acz} which summarises how the various corrections to the scalar potential scale.}
    \label{tab:junghans}
\end{table}

Notice that among the corrections in particular, the $\alpha'$ ones arise at the same subleading level as the flux corrections and this is one of the motivations of \cite{Andriot:2023fss} to consider more generic expansions where a decoupling between the two can occur to address the question of robustness of the mass spectrum in terms of flux corrections only.

As before, these expansions with the smeared solutions as the leading contribution and its subleading corrections can only be trusted not too close from the O-planes at substringy distances (decreasing with $n$) since the procedure of course does not resolve the O-plane singularities from a stringy point of view\footnote{Note that the behaviour near the O-planes is expected by \cite{Junghans:2020acz} to be different from the supergravity smoothing attempt of \cite{Saracco:2012wc,Saracco:2013aoa} discussed earlier but the authors comment that the no-go of \cite{Gautason:2015tig} against such a smoothing should not apply to DGKT and therefore does not rule out this possibility.}. At the supergravity level it is however argued that what happens near the O-planes should not spoil the construction and the validity of the smeared approximation almost everywhere in the compact space at large $n$. As an important remark, reference \cite{Junghans:2020acz} explains that it is theoretically possible (and not checked at the moment) that the corrections to the scalar potential, even if subleading, could induce a breaking of supersymmetry. This would be bad regarding potential instabilities  (see sect.~\ref{sec:unstab}). Still, the would-be breaking would disappear at large $n$.

The precise geometry of the internal space for this extension of the smeared solution is expected to be described by a $SU(3)\times SU(3)$ structure \cite{Acharya:2006ne,McOrist:2012yc,Saracco:2012wc} and this is precisely what has been investigated in \cite{Marchesano:2020qvg} in the supersymmetric case where a first-order solution with localised sources has been built by focusing on the supersymmetry equation. The method is different from the strategy outlined above but both of them reach the same conclusion that corrections to the smeared approximation are suppressed everywhere in the internal space at large $n$ except near the O-planes. Extension of this $SU(3)\times SU(3)$-structure formalism to encompass non-supersymmetric scale-separated vacua have then been developed in \cite{Marchesano:2021ycx,Marchesano:2022rpr}. This approach actually directly addresses the third point raised in \ref{sec:against_smearing} and shows that it is not an obstacle.

One can then expect similar arguments to hold also for the massless T-dual constructions presented earlier. This has been explicitly shown for the massless scale-separated vacua of \cite{Cribiori:2021djm} built from T-duality in a toroidal setup. The authors show that the smeared approximation provides a correct asymptotic description of the true localised solution, at least at first order. In addition, they also claim to give evidence for a consistent M-theory lift with scale separation, which contrasts with sect.~\ref{sec:Mlift}. The success is argued to come from more general flux scalings and a proper consideration of the O-planes backreaction to solve the 11$d$ equations of motion. Moreover the M-theory equations seem to be sourceless, in tension with the no-go presented in sect.~\ref{sec:unavoidable}. This suggests that the geometry actually evades some assumptions of the theorem and could thus provide a counterexample to the conjecture formulated in \cite{Collins:2022nux} by hinting at an internal geometry for which the KK scale can be decoupled from the curvature. However higher-order corrections are expected to account for intersecting O-planes in the covering space and could have strong consequences on the backreaction computation.

Finally, notice that same backreaction computation techniques have been applied to investigate the scale-separated AdS$_3$ vacua mentioned earlier \cite{Emelin:2021gzx}.

\subsubsection{Source intersection}
\label{sec:intersection}

As we have seen several times, the fact that orientifold planes intersect (from the point of view of the covering space) is a source of concern for the robustness of the backreaction analyses. In contrast with these worries, a very recent study \cite{Junghans:2023yue} showed that depending on the compactification space at hand, one may actually not produce any intersection of O6-planes in the covering space. Such intersections are argued to arise typically in orbifold setups but to be sometimes absent on smooth Calabi--Yau manifolds and in particular on resolved orbifolds. Without intersection loci in the covering space, one would therefore expect no pathology to show up and spoil the backreaction analysis. The orbifold limit where potential intersections reappear could then in principle be reconstructed from the resolved setup and would then feature no pathology either. If correct, this provides another non-trivial consistency check for scale-separated vacua by opening the possibility to evade the presence of source intersections in the covering space and their associated concerns. As a consequence, one would expect the localisation procedures described above to be perfectly fine in that regard.\\

All this brings strong evidence that the smearing approximation makes sense to give an accurate description of a true localised solution at large $n$. Of course this is no proof that fully-fledged uplifts exist and one can worry that fatal things happen at higher-order in the expansion but it is important to realise that things could already have gone wrong at this point and it is non-trivial that they did not. What happens near the O-planes also remains unknown and for this M-theory uplifts should be investigated further. For massive type IIA solutions, the Romans mass prevents the construction of a M-theory lift \cite{Aharony:2010af} such that the massless T-dual constructions \cite{Cribiori:2021djm,Carrasco:2023hta} seem more promising from this point of view, as explicitly shown in \cite{Cribiori:2021djm}.

\subsubsection{Does scale separation survive the lift?}

Before concluding this section notice that another concern when uplifting, regardless of the existence or not of a consistent 10$d$ solution, is to ensure that the properties of the putative uplift are not modified too much compared to the smeared approximation such that scale separation remains a feature of the uplifted solution. For example, it has been shown that some solutions with metric fluxes for which a 10$d$ lift is known \cite{Aldazabal:2007sn} lose their scale-separation property when properly analysed from the 10$d$ point of view \cite{Font:2019uva} due to corrections to the KK scale. Indeed in this setup, a proper 10$d$ computation of the dependence of the KK scale on the geometric fluxes show that they cancel out the naive 4$d$ result and scale separation disappears.

\subsection{Conclusion: So far so good?}

We saw that despite generic arguments made against them due to the smearing approximation, the scale-separated families of vacua built from a 4$d$ perspective successfully overcome the criticisms regarding their 10$d$ completion. Indeed, things could have fatally failed already at this point but as a matter of fact it is not the case: In the best case scenarios, the scale-separated families of vacua built from a low-dimensional perspective are robust against uplifting, the smeared approximation indeed arises as a dominant contribution from an expansion when the O6-planes are localised in the 10$d$ equations of motion, intersection of sources can be absent, scale separation remains a 10$d$ feature and eventually evidence for the existence of an M-theory lift can be given, at least in the massless case.

However, even with an optimistic view of the uplifting problem thanks to the literature we reviewed here, other considerations may suggest that the scale-separated families of vacua are still inconsistent in the end. First of all, the Swampland Program \cite{Vafa:2005ui,Brennan:2017rbf,Palti:2019pca,vanBeest:2021lhn,Grana:2021zvf} has come up with a variety of conjectures, some of them specific to AdS vacua, that could threaten the possibility to achieve scale separation. On the other side of the same coin, efforts have been developed to try to uncover the holographic dual of the DGKT proposal and more generally of scale-separated bulk spacetimes, with no positive result so far. We propose to review this relationship between swampland conjectures and scale separation as well as holographic considerations in the next section.

\section{Swampland conjectures, holography and scale separation}
\label{sec:conjs}

\begin{figure}[ht!]
    \centering
    \includegraphics[scale=0.3]{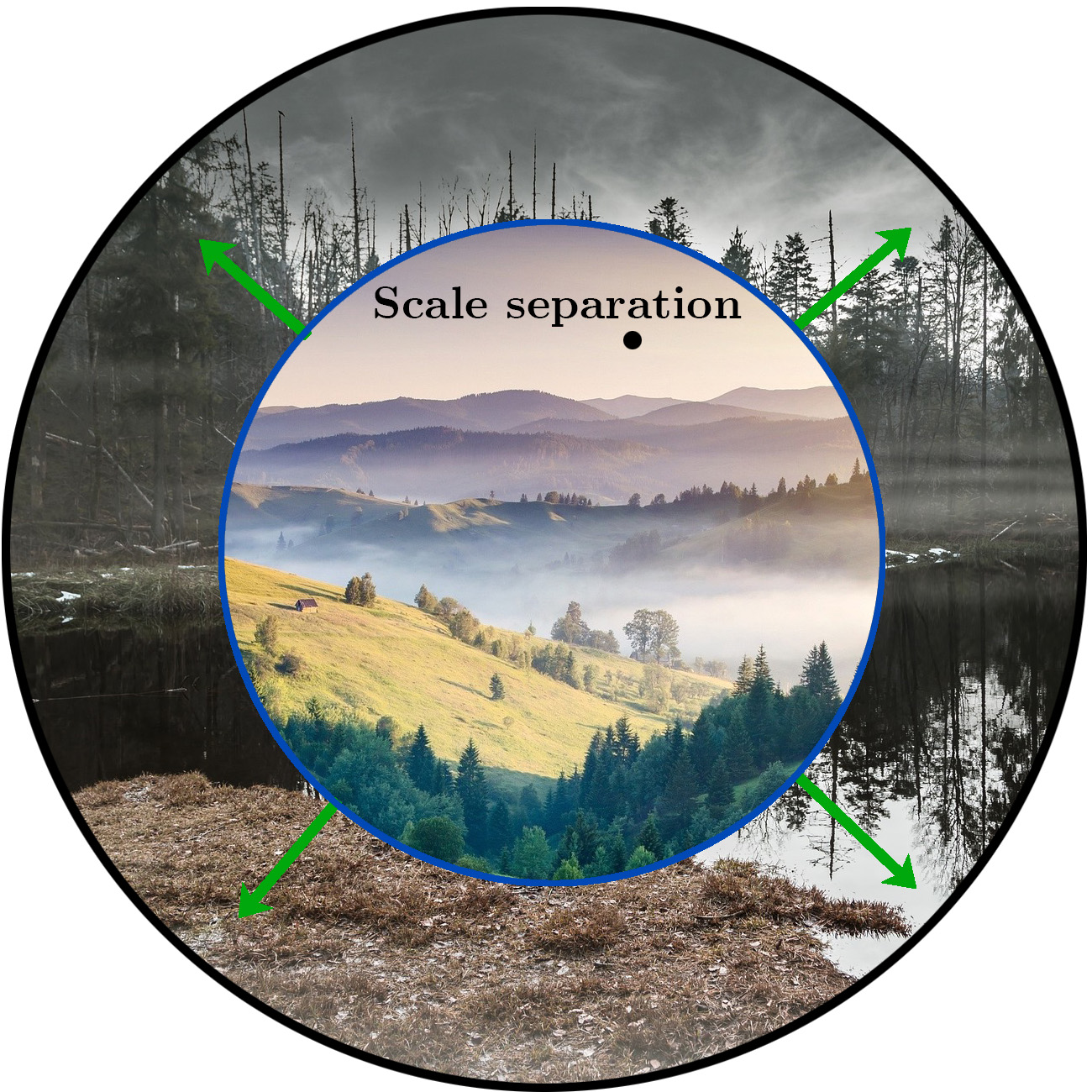}
    \qquad
    \includegraphics[scale=0.3]{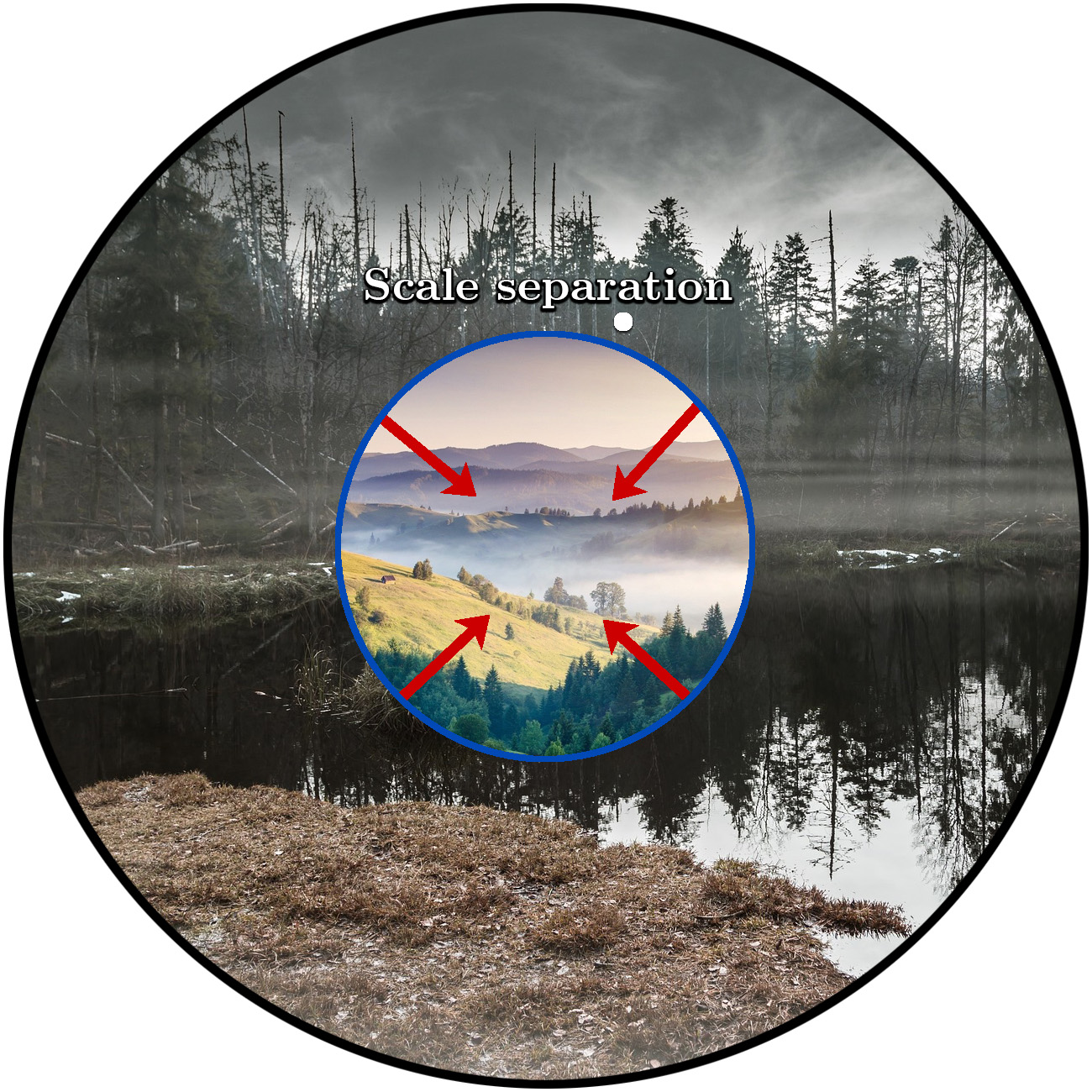}
    \caption{From a bottom-up perspective on the left, scale separation seems to be part of the Landscape in contrast with a top-down approach where some swampland conjectures suggest it is actually in the Swampland.}
    \label{fig:clash}
\end{figure}

Bottom-up constructions of phenomenologically appealing vacua in string theory (in our case: Vacua featuring parametric scale separation) are crucial steps for our quest to link our universe to the theory. The bottom-up nature of such scenarios of course comes with a number of caveats related to whether the results can really be obtained from a fully-fledged string theory embedding as we have seen, but one has to start somewhere. From a Swampland perspective, these constructions contribute to build the Landscape from inside by pushing its boundaries outwards. The Swampland Program philosophy \cite{Vafa:2005ui,Brennan:2017rbf,Palti:2019pca,vanBeest:2021lhn,Grana:2021zvf} tackles the same question of delineating the Landscape from the Swampland but from the opposite direction: By pushing the Landscape boundaries inwards thanks to statements that act like no-go for string theory and more generally quantum gravity\footnote{Due to the ambition of the Swampland Program to provide universal patterns about what is consistent with quantum gravity and what is not, we will sometimes refer to this approach as being ``top-down'', precisely in this sense.}. Both approaches are valuable and contribute to a better understanding of the string Landscape. For parametric scale separation (similarly to other properties), the bottom-up approach clashes with the Swampland perspective since the former suggests it is achievable while some swampland conjectures are in tension with it (see fig.~\ref{fig:clash} for a schematic illustration of this clash and a pictorial representation of the two approaches). Whether parametric scale separation is in the Landscape or not and under which conditions is thus still an open question despite a lot of progress made on the two sides of the borderline. So far in this review we were inside the Landscape pushing the boundaries outwards but we will now switch gears, jump outside and compress them.

The purpose of this section is then to review how scale separation has been investigated through the Swampland lens in the literature. We first see that the No Global Symmetry Conjecture and the Weak Gravity Conjecture, ones of the most well-established conjectures of all, have been used to rule out scale separation with extended supersymmetry. We then mention the strongest stab at scale separation from the Swampland Program, namely the Strong AdS Distance Conjecture before reviewing softer versions of the statement and how they could accommodate for scale separation. Eventually, we mention the AdS instability conjecture, which in particular casts doubts on the robustness of the non-supersymmetric vacua exhibiting scale separation. We eventually end the section with a review of the status of scale separation from an holographic point of view.

\subsection{Ruling out extended supersymmetry}

As we will review now, recent works suggest that in light of the most robust swampland conjectures, extended supersymmetry in 4$d$ (or more generally having too many supercharges) is incompatible with scale separation. As a result, this leaves only non-supersymmetric $\N=0$ or minimally supersymmetric $\N=1$, $d\leq 4$ vacua outside from the Swampland in that respect. Consistently, all the scale-separated scenarios that we reviewed in the previous sections fall into this class of vacua.

\subsubsection{The weak gravity conjecture}

The Weak Gravity Conjecture \cite{ArkaniHamed:2006dz} in its magnetic form asserts that the EFT cutoff $\LUV$ in a gauge theory with coupling $g$ for a $p$-form gauge fields is bounded by
\begin{equation}
\LUV\leq \left(g^2\Mp^{d-2}\right)^{\frac{1}{2(p+1)}}\, .   
\end{equation}
This conjecture was used in \cite{Cribiori:2022trc} to study 4$d$ AdS vacua with $\N\geq 2$ in the presence of an unbroken $U(1)$ gauge factor to apply the conjecture. In this context, the authors derive bounds of the form
\begin{equation}
\label{eq:LUV}
\Lambda\gtrsim q^2\LUV^2\, ,
\end{equation}
where $q$ is some non-vanishing charge, thanks to the conjectured bound on the coupling constant. In agreement with the Emergent String Conjecture \cite{Lee:2019oct}, if one identifies $\LUV$ with the KK scale $\MK$, then \eqref{eq:LUV} implies
\begin{equation}
\Rads\MK\lesssim 1\, ,
\end{equation}
such that the scale separation condition \eqref{eq:scale_sep} is unachievable. In the same spirit, five-dimensional AdS vacua have been investigated in light of the Weak Gravity Conjecture in \cite{Cribiori:2023ihv} where similar bounds like \eqref{eq:LUV} have been found\footnote{In this context the Weak Gravity Conjecture is actually shown to imply the Strong AdS Distance Conjecture that we will mention soon }. The bounds are derived for minimal $\N=2$ supersymmetry in 5$d$ ($8$ supercharges) and maximal $\N=8$ ($32$ supercharges) and are expected to hold for intermediate amounts of supercharges as well as for higher dimensions. If true, these arguments lead us to conclude that the only corner of the Landscape where AdS scale separation can possibly be achieved corresponds to $d\leq 4$ and $\N\leq 1$ vacua.

\subsubsection{No Global Symmetries and extreme scale separation}

The direct use of the Weak Gravity Conjecture for AdS vacua has been argued to be a strong assumption \cite{Montero:2022ghl} and it was proposed there to analyse scale separation in the case where the internal space is absent or frozen at Planck length such that no tower of light states arises. This is dubbed \emph{extreme} scale separation and is argued to be in the Swampland due to the No Global Symmetry Conjecture \cite{Banks:1988yz,Banks:2010zn}. The argument goes as follows: In such theories the $SU(4)$ R-symmetry is gauged and if one parametrically scales the AdS radius and send it to infinity, the R-symmetry becomes global while the Planck scale remains finite\footnote{Note that invoking the R-symmetry as an obstacle to scale separation has also been explored in \cite{Polchinski:2009ch} from an holographic point of view (see sect.~\ref{sec:holo}).}. This result then also goes in the direction that too many supersymmetries are bad for scale separation.

\subsection{The Strong AdS Distance Conjecture}

The strong version of the AdS Distance Conjecture \cite{Lust:2019zwm} is the most direct stab at scale separation from the Swampland Program since it simply rules it out for supersymmetric vacua in all circumstances. It states that for any AdS vacuum, there is an infinite tower of states that become light in the limit where the cosmological constant goes to zero with mass
\begin{equation}
\frac{m}{\Mp}\sim |\Lambda|^\alpha\, ,    
\end{equation}
where $\alpha$ is an order one constant. The strong part of the conjecture then further says that $\alpha=\half$ for supersymmetric vacua. If the mass of the tower is identified with the KK scale $\MK$, one straightforwardly deduce that $\Rads\MK\lesssim 1$ and scale separation is forbidden. All the scale-separated constructions we have reviewed in the previous section are thus in direct tension with this strong version of the conjecture.

As we will see now, a weaker form of the statement or some refinements can then accommodate for scale separation.

\subsection{Weak AdS Distance Conjecture}
\label{sec:weak}

In its weak form, the conjecture simply relaxes the sharp condition $\alpha=\half$ for supersymmetric vacua, which is precisely at the boundary for forbidding scale separation. Any smaller value for $\alpha$ is then compatible with scale separation and all the scale-separated constructions satisfy this weaker version of the conjecture. For the DGKT scenario, the scalings \eqref{eq:scaling_MKK} and \eqref{eq:scaling_Lambda} yield $\alpha=\frac{7}{18}$. Note that for the scale separation mechanism involving Casimir energy in M-theory described in sect.~\ref{sec:casimir}, the scaling is smaller and given by $\alpha=\frac{1}{4}$. Such a small value turns out to saturate the recently proposed lower bound $1/d$ for this parameter \cite{Rudelius:2021oaz,Castellano:2021mmx,Montero:2022prj}.

In reference \cite{Junghans:2020acz} an argument is given to try to explain why the DGKT-like vacua evade the strong version of the conjecture even if it is natural to think that it holds for most AdS vacua. The idea is the following: Using rescaling properties of the supergravity equations of motion without O-planes, which involve two parameters \cite{Witten:1985xb,Burgess:1985zz}, one can evaluate the behaviour of the KK scale and the cosmological constant in the limit of large volume and weak coupling to conclude that in this case indeed $\alpha=\half$. If the two rescaling parameters are all there is to engineer a controlled regime, then the strong version of the conjecture is recovered in this case. The addition of O-planes breaks at least one of the two rescaling symmetries and when one survives, it cannot be used to go to weak coupling. However, the crucial property of the DGKT-like vacua is that they feature this new scaling symmetry linked to the unbounded $G_4$ flux $n$ thanks to vanishing terms in the equations of motion and it can then be used to achieve scale separation as we have seen.

\subsection{Refined AdS Distance Conjecture}

Another refinement of the AdS Distance Conjecture has been proposed in \cite{Buratti:2020kda} and could also allow for scale-separation in DGKT-like vacua. The new statement is called $\Z_n$ Refined Strong AdS Conjecture and implies that when a discrete $\Z_n$ symmetry for domain walls is present with $n$ large, then the tower of light particles arising when the cosmological constant approaches zero (while $\Lambda n$ also goes to zero) behaves like
\begin{equation}
\label{eq:refined}
\frac{m}{\Mp}\sim |\Lambda|^\half\sqrt{n}\, .
\end{equation}
Such a non-trivial discrete three-form $\Z_n$ symmetry acting on domain walls interpolating between different flux values is shown to exist in the DGKT vacua (with $n$ matching the scaling parameter) and the tower becoming light as $\Lambda$ goes to zero then follows the refined behaviour \eqref{eq:refined}, in agreement with the scaling \eqref{eq:sep_DGKT} of scale separation in DGKT. This refinement saves 4$d$ DGKT-like vacua but it has been remarked \cite{Apers:2022zjx} that the required discrete symmetries are not present for the AdS$_3$ scale-separated constructions mentioned earlier in sect.~\ref{sec:3d} such that these vacua even violate this refined conjecture. On the other hand, the $\Z_n$ symmetry is expected to be invariant under dualities such that the massless type IIA models that are T-duals of DGKT satisfy the refined conjecture as well.\\

Notice also for completeness that yet another weaker version of the Strong AdS Distance Conjecture has been formulated \cite{Moritz:2017xto} and dubbed AdS/Moduli Scale Separation Conjecture \cite{Blumenhagen:2019vgj}. It states that the AdS radius cannot be decoupled at least from the lightest modulus with mass $m_\phi$:
\begin{equation}
\Rads m_\phi\lesssim 1\, ,
\end{equation}
and has been argued to be an obstacle to de Sitter uplifting since it forbids too steep AdS vacua that are necessary not to destroy the minimum after uplifting.

\subsection{Swampland Distance Conjecture}

The AdS Distance Conjecture can be thought to be the application of the Swampland Distance Conjecture \cite{Ooguri:2006in}, which states that at infinite distance in moduli space one should encounter an exponentially light tower of fields, to the space of spacetime metrics (see \cite{Li:2023gtt} for an attempt to derive a well-defined distance in this setup). Forgetting about the AdS distance and focusing on the more established standard notion of distance in moduli space, a scalar field has been proposed in \cite{Shiu:2022oti} to interpolate between scale-separated DGKT vacua with different flux numbers (in a similar way as in \cite{Kachru:2002gs,Scalisi:2020jal} in type IIB and see also \cite{Basile:2023rvm} for a similar discussion) and has been shown to satisfy the Swampland Distance Conjecture. It describes D4-branes of codimension one in the internal space which discharge the flux number as they convert the whole internal space. This provides further evidence that scale-separated solutions might not be in the Swampland and that the Strong AdS Distance Conjecture should not hold in full generality. The same kind of arguments are given for the AdS$_3$ vacua built in \cite{Farakos:2023wps} where it is shown that the Swampland Distance Conjecture holds when interpolating between vacua with full parametric scale separation between AdS$_3$ and the whole $G_2$ manifold at large $n$ as well as vacua where an additional $S^1$ is large to produce scale separation in 4$d$ at moderate $n$. The authors also connect their constructions with the Strong Spin-2 Conjecture of \cite{Klaewer:2018yxi}, which argues for a universal cutoff that depends on the lightest spin-2 field mass\footnote{This conjecture has also been discussed in \cite{DeLuca:2021ojx} which investigates spin-2 fields in specific compactifications.}. The strong version asserts it also applies to massive gravity where no massless graviton is present.

Fundamentally, because the interpolation mechanism involves massive fields, what is probed by these results is a stronger version of the Swampland Distance Conjecture and not the original one where the trajectory is inside the moduli space. This is also the case for the very recent mechanism proposed in \cite{Shiu:2023bay} to interpolate between vacua with different background fluxes $F_k$. The schematic idea is to describe a domain wall formed by a D($8-k$)-brane thanks to a one-dimensional higher spacetime-filling (non-BPS) D($9-k$)-brane which contains a tachyon $T$. The two tachyon vacua at $T=\pm\infty$ then have different units of flux and one can evaluate the distance travelled by the tachyon to connect the vacua. This distance then agrees with the Swampland Distance Conjecture applied to the KK tower, but with a small $\alpha$ parameter controlling the speed of the decay in tension with the sharpened version of the Swampland Distance Conjecture \cite{Etheredge:2022opl}. Overall this mechanism with the tachyonic scalar is more model independent than the construction described above with the scalar being some brane position. As such it has been applied in various setups and in particular in the massless type IIA scale-separated solutions.

\subsection{AdS Instability Conjecture}
\label{sec:unstab}

The AdS Instability Conjecture \cite{Ooguri:2016pdq,Freivogel:2016qwc} states that non-supersymmetric AdS vacua are unstable and therefore should have at least one non-perturbative decay channel available. If true, this of course puts in particular all the non-supersymmetric scale-separated solutions at the verge of the Swampland since they cannot last forever\footnote{From a phenomenological point of view one could well be happy with a sufficiently long-lived scale-separated vacuum and not be too bothered by these non-perturbative instability analyses.}. A series of works have been precisely dedicated to tackle this issue and evaluate the decay channels available to non-supersymmetric AdS$_4$ orientifold vacua. Note that if the couple formed by this conjecture together with the Strong AdS Distance Conjecture is correct, then stable scale separation is in the Swampland in all circumstances.

\subsubsection[4$d$ membranes]{\bm 4$d$ membranes}

From the 4$d$ EFT point of view the membranes that can nucleate and trigger a non-perturbative instability are made of D4-branes wrapping a two-cycle or made of D8-branes wrapping the whole internal space \cite{Lanza:2019xxg,Lanza:2020qmt}. To investigate stability one needs to evaluate the tensions and charges of all possible membranes and conclude that a decay channel is open whenever the charge $Q$ of such membrane is bigger than its tension $T$ (which is precisely what the sharpened version of the Weak Gravity Conjecture (sWGC) \cite{Ooguri:2016pdq} states) whereas it is only marginal if the membrane saturates the BPS bound $Q=T$. In \cite{Narayan:2010em} the membranes arising from D4-branes wrapping two-cycles have been analysed in the smeared approximation for some branches of AdS$_4$ solutions and shown to saturate the BPS bound even in non-supersymmetric vacua, thus triggering no instability at this level of approximation. However whenever a membrane is marginal one should worry about further corrections that could destroy marginality and push the charge to tension ratio in one direction or the other.

In \cite{Marchesano:2021ycx} the authors made use of the first-order formalism beyond the smearing approximation developed in \cite{Marchesano:2020qvg} to precisely check that even at this next-to-leading order the D4-branes wrapping two-cycles have $Q_{\rm D4}=T_{\rm D4}$. They also explored a potential decay channel through nucleation of D8-branes in BIonic configurations that require also the presence of spacetime-filling D6-branes. Beyond the leading order where the membranes saturate the BPS bound, superextremality $Q_{\rm D8}>T_{\rm D8}$ has been shown for a simple toroidal example but a further analysis of these membranes and the corrections they received performed in \cite{Casas:2022mnz} suggested that it is actually possible to engineer vacua where $Q_{\rm D8}<T_{\rm D8}$ and therefore violate the Weak Gravity Conjecture for membranes. This statement was revised in \cite{Marchesano:2022rpr} where more intricate D8-brane systems were shown to give rise to superextremal 4$d$ membranes, saving the conjecture.

A nice summary of these considerations is provided in \cite[Table 3]{Marchesano:2022rpr} where results for one supersymmetric branch of solutions and two non-supersymmetric ones are displayed. We copy it here in table~\ref{tab:membranes}.

\begin{table}[!ht]
    \centering
    \begin{tabular}{|c||c|c|c|c|c|}
    \hline
    Branch & SUSY & pert. stable & sWGC D4 & sWGC D8 & non-pert. stable\\\hline\hline
    {\bf A1-S1+} & Yes & Yes & Yes & Yes & Yes\\\hline
    {\bf A1-S1-} & No & Yes & Marginal & Yes & unclear if $N_{\rm D6}=0$\\\hline
    {\bf A2-S1$\pm$} & No & Yes & Yes & Yes & No\\\hline
    \end{tabular}
    \caption{Table extracted from \cite{Marchesano:2022rpr} which summarises the status of the sWGC and stability of the vacua in various branches of solutions uncovered in \cite{Marchesano:2019hfb}. The ``A1-S1'' branch is the one containing the DGKT-like vacua.}
    \label{tab:membranes}
\end{table}

Both from D4- and D8-branes, the membranes in the supersymmetric vacua (first line) saturate the BPS bound and the construction is then non-perturbatively stable. For one of the non-supersymmetric branch (third line), the sharpened Weak Gravity Conjecture is satisfied for both channels and it is thus unstable. For the other branch however (second line), which is the one containing the DGKT-like constructions, marginality remains for the D4-branes whereas the sharpened conjecture is satisfied for the D8-branes. But it is important to realise that the decay channel through BIonic systems of D8-branes require D6-branes and thus naively the sharpened Weak Gravity Conjecture does not hold if such branes are not present. Not having D6-branes is precisely what happens in the DGKT construction where the tadpole is saturated with fluxes. In this setup, the apparent absence of a D8-brane decay channel together with marginality for the D4-branes is then a glimpse of hope for the existence and stability of non-supersymmetric scale-separated vacua. The authors also point at the fact that these vacua for which non-perturbative stability is unclear are precisely those that feature strange properties from an holographic point of view (integer conformal dimensions), which will be discussed in sect~\ref{sec:holo}.

\subsubsection{Non-perturbative supersymmetry breaking?}
\label{sec:non-pert}

The supersymmetric DGKT vacua have been suggested to actually admit non-supersymmetric domain walls coming from D4-branes \cite{MiguelInProgress}. Due to quantum corrections, such domain walls are shown to be subextremal with $T_{\rm D4}>Q_{\rm D4}$ instead of satisfying $T_{\rm D4}=Q_{\rm D4}$ as one would expect for a supersymmetric setup. This results is then in tension with the membrane Weak Gravity Conjecture and one could interpret this tension in two ways. Either by arguing that the Weak Gravity Conjecture for membranes is not fully correct, in agreement with the fact that it is not as well-grounded as the original particle version of the conjecture. Or one could conclude that something is wrong in DGKT and it should suffer from some pathology. To argue in this direction, the authors checked that these subextremal membranes arise only in DGKT-like vacua and not in other 4$d$ $\N=1$ AdS vacua whose UV completion is clearer. This then raises suspicions on DGKT. On the other hand, the authors are also considering the possibility that despite such non-supersymmetric domain walls, the whole setup could actually stay supersymmetric.

One can wonder how such effects could affect the massless T-dual constructions. But since T-duality is not guaranteed to hold at the non-perturbative level \cite{Aspinwall:1999ii}, if a non-perturbative supersymmetry breaking is found in DGKT, then one cannot directly transpose the result and a similar analysis should be redone again in the T-dual frame.

\subsubsection{Decay to nothing}

As an aside remark, note that bubbles of nothing \cite{Witten:1981gj,Brill:1991qe,Fabinger:2000jd,Horowitz:2007pr,Blanco-Pillado:2010xww,Brown:2010mf,Brown:2011gt,Blanco-Pillado:2010vdp,Blanco-Pillado:2016xvf,Ooguri:2017njy,Acharya:2019mcu,Dibitetto:2020csn,Draper:2021qtc,Draper:2021ujg,Draper:2023ulp,Friedrich:2023tid} have been suggested to provide a universal decay channel for every non-supersymmetric vacuum \cite{GarciaEtxebarria:2020xsr}. This universality takes root into the Cobordism Conjecture \cite{McNamara:2019rup} that states that in a quantum theory of gravity compactified on a $d$-dimensional manifold the cobordism classes should vanish: $\Omega^{\rm QG}_d=0$. This statement translates into the possibility to always allow for a bubble of nothing at least from the topological point of view. If in addition the nucleation is always dynamically favourable then this provides a universal decay process. Of course the latter condition is crucial and not proven up to date.

\subsection{Summary}

Table~\ref{tab:conjs} provides a visual summary of the status of the Swampland Conjectures in the supersymmetric, classical and parametrically scale-separated constructions that we have reviewed so far.

\begin{table}[!ht]
    \centering
    \begin{tabular}{|c|c|c|c|c|}
    \hline
    \diagbox{Model (SUSY)}{Conjecture} &  SADC & WADC & $\Z_n$ RSADC & SDC\\\hline\hline
    DGKT-like & \includegraphics[height=0.5cm]{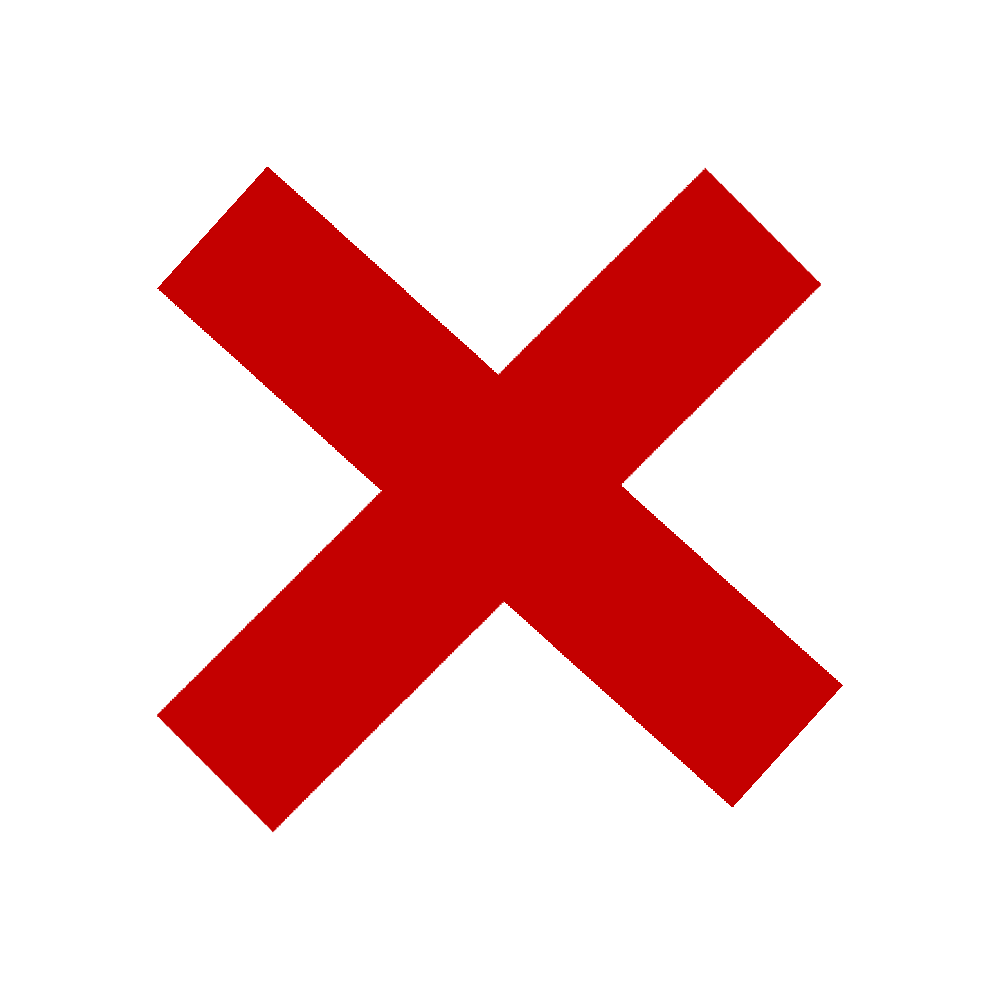} & \includegraphics[height=0.5cm]{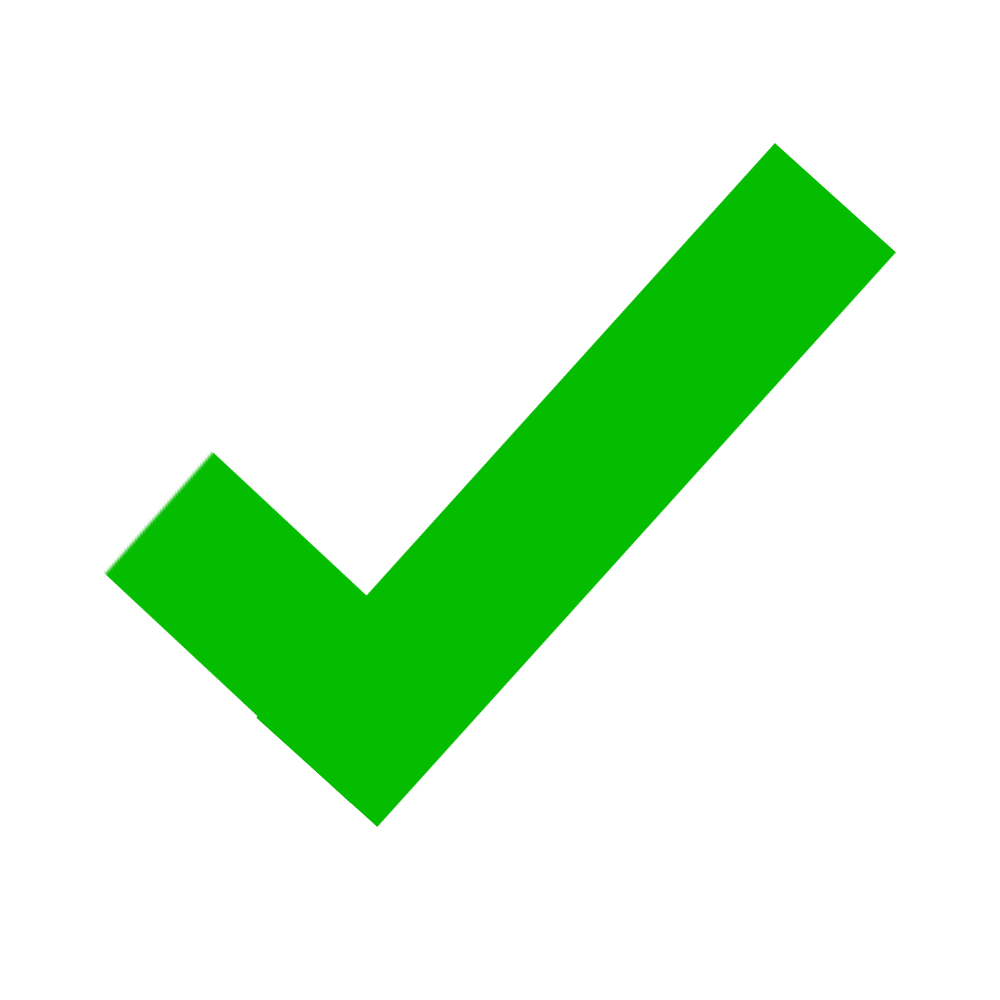} & \includegraphics[height=0.5cm]{tick.png} & \includegraphics[height=0.5cm]{tick.png}\\\hline
    Massless T-duals & \includegraphics[height=0.5cm]{cross.png} & \includegraphics[height=0.5cm]{tick.png} & \includegraphics[height=0.5cm]{tick.png} & \includegraphics[height=0.5cm]{tick.png}\\\hline
    AdS$_3$ vacua & \includegraphics[height=0.5cm]{cross.png} & \includegraphics[height=0.5cm]{tick.png} & \includegraphics[height=0.5cm]{cross.png} & \includegraphics[height=0.5cm]{tick.png} \\ \hline
    \end{tabular}
    \caption{This table summarises the status of various conjectures in the three mains class of supersymmetric scale-separated vacua that we have encountered: The DGKT-like solutions, their massless type IIA double T-duals and the AdS$_3$ vacua. The acronyms are as follows. SADC: Strong Anti de Sitter Distance Conjecture. WADC: Weak Anti de Sitter Distance Conjecture. $\Z_n$ RSADC: Refined Strong Anti de Sitter Distance Conjecture. SDC: Swampland Distance Conjecture.}
    \label{tab:conjs}
\end{table}

\subsection{Holographic considerations}
\label{sec:holo}

Holography could either be salvation or doom. Salvation if a proper CFT dual of DGKT-like vacua could be found which would settle the debate about their consistency and stability, or doom if we realise that it would require specific properties of the CFT that are impossible to achieve. The recent developments in the literature are unfortunately pointing more towards the latter case since no CFT dual has been found yet and properties that these would-be duals should have to reflect scale-separation in the bulk are rather peculiar and not encountered in common examples. Indeed the most famous AdS/CFT pairs like AdS$_5\times S^5$ in type IIB or AdS$_4\times S^7$ in M-theory have an internal sphere of order-one size in AdS radius units. In the AdS$_5\times S^5$ case, the sphere is threaded by $n$ units of $F_5$ flux, dual to the super Yang--Mills gauge group. This flux sets the size of the sphere $R_{S^5}$ as well as the AdS scale like
\begin{equation}
\Rads\Mp\sim R_{S^5}\Mp\sim n^{\frac{2}{3}}\, ,
\end{equation}
where $\Mp$ denotes here the five-dimensional Planck scale and from which we read that no separation of scales can be achieved due to the identical scalings.

In addition to these well-controlled setups, more specific works suggest that CFT's associated with scale-separated bulk spacetimes simply cannot exist. From a broader point of view, applying a Swampland mindset in the context of holography and trying to translate the various conjectures into the CFT language has been a thriving topic in the recent years \cite{Harlow:2015lma,Giombi:2017mxl,Harlow:2018jwu,Harlow:2018tng,Montero:2018fns,Baume:2020dqd,Perlmutter:2020buo,Harlow:2020bee,Aharony:2021mpc,Baume:2023msm}. Notice that the following review of holographic considerations does no much better than what is nicely summarised in \cite[section 5.6]{VanRiet:2023pnx}.

\subsubsection{Large gap}

We know from the holographic dictionary \cite{Maldacena:1997re} that the conformal dimension $\Delta$ dual to a bulk scalar of mass $m$ in AdS$_d$/CFT$_{d-1}$ is given by \cite{Witten:1998qj,Gubser:1998bc,Aharony:1999ti}
\begin{equation}
\label{eq:Delta}
\Delta=\frac{d-1}{2}+\half\sqrt{(d-1)^2+4m^2\Rads^2}\, ,
\end{equation}
and similar behaviours with respect to the mass and the AdS radius hold for other kinds of fields \cite{Aharony:1999ti,Freedman:1999gp}. Some generic properties of the CFT duals and others specific to scale-separated bulk spacetimes can then easily be derived from this relation. First of all, requiring in the bulk $\Rads\gg \ell_s$ to have a local description below the AdS scale implies that the higher-spin excited states with mass $\sim \frac{1}{\ell_s}$ have a large conformal dimension \cite{Heemskerk:2009pn}. There is thus a large difference in conformal dimensions between the light content of the theory and the higher-spin states. One sometimes refers to this difference as a ``gap'' in the conformal dimensions even if strictly speaking there is no gap at this stage because of the KK modes that fill the difference in between in non scale-separated setups. Much more details about properties of CFT's with holographic duals can be found in \cite{El-Showk:2011yvt,Fitzpatrick:2012cg}.

In addition, scale-separated solutions also have a decoupling in conformal dimensions between only a few low-lying states and their associated KK towers which are very massive and have large conformal dimension as well due to the scale separation condition $\Rads\MK\gg 1$ \cite{Polchinski:2009ch,deAlwis:2014wia,Conlon:2018vov,Conlon:2020wmc}. There is in this case a proper gap of conformal dimensions between the massless states and the first massive KK mode and higher-spin states. The desired hierarchies and their consequences on the conformal gap are nicely summarised in the following relations that hold for setups without and with scale separation:
\begin{equation}
\begin{aligned}
&\text{No scale separation: }&&\underbrace{\Rads\sim L_{\rm KK}\equiv\frac{1}{\MK}\gg\ell_s}_{\text{Higher-spin ``gap''}}\, ,\\
&\text{Scale separation: }&&\underbrace{\Rads\gg L_{\rm KK}}_{\text{KK tower gap}}\gtrsim \ell_s.
\end{aligned}
\end{equation}
In the end, such a low number of $\mathcal{O}(1)$ conformal dimension operators is not usual in known CFT's. Note that such large gaps are often interpreted as having a strongly coupled CFT \cite{Polchinski:2009ch}.

\subsubsection{More specific results}

Besides these basic generalities, more specific investigations of the properties of CFT's dual to scale-separated bulk spacetimes have been performed in the literature, some of them we have already mentioned in this review as no-go's for scale separation without emphasising their holographic origin.

First we already briefly pointed at \cite{Polchinski:2009ch} which argues that the presence of an R-symmetry can protect operators dual to KK excitations and thus prevent their conformal dimensions to become parametrically large. A $d=4$, $\N=1$ theory however (like the DGKT-like vacua) remains unprotected since there is no R-symmetry left. To get a scale-separated bulk spacetime, it was thus conjectured in the paper that what is needed is the combination of strong coupling due to the large gap together with the absence of R-symmetry. In the same spirit, we also already mentioned the reference \cite{Montero:2022ghl}. It is argued there that in case of extreme scale separation (meaning $\alpha=0$ in the AdS Distance Conjecture) with too many supercharges, the gauged R-symmetry is expected to become global and then put these cases into the Swampland. On the other hand, the AdS Instability Conjecture which forbids non-supersymmetric AdS vacua can be motivated with holography by arguing that a dual description cannot exist due to an instantaneous decay near the boundary \cite{Horowitz:2007pr,Harlow:2010az,Ooguri:2016pdq}.

In sect.~\ref{sec:localisation} we also already talked about the reference \cite{Collins:2022nux} and its conjecture in tension with the M-theory geometry of the toroidal T-dual of DGKT proposed in \cite{Cribiori:2021djm}. More precisely, the conjecture is rooted in holography and asserts that there is a universal upper bound on the conformal dimension of the first non-trivial spin-two operator for any CFT. This translates into a lower bound on the diameter of the internal space in AdS radius units which then rules out parametric scale separation. The conjecture is motivated by the study of CFT's dual to gravitational theories with internal spaces that are Sasaki--Einstein manifolds coming from branes that probe regular or Fermat type Calabi--Yau singularities. In all cases an upper bound on the lowest spin-2 conformal dimension depending only on the number of dimensions of the internal space is found.

Another holographic argument takes a stab at the scale-separated AdS vacua used as first steps in the KKLT \cite{Kachru:2003aw} scenario. Properties of the CFT's dual to KKLT have been investigated in \cite{deAlwis:2014wia} and more recently in \cite{Lust:2022lfc} where it has been argued that it is not consistent to get a CFT dual to the AdS first step. Indeed in this setup, an upper bound on the central charge (rooted in the tadpole bound) is proposed which in turns translate into an upper bound on the AdS radius. This upper bound then forbids exponentially small cosmological constants to arise. In addition, even for the most scale-separated solutions, the AdS radius is shown to be of the order of the cutoff length scale, putting theses KKLT AdS out of reach of the EFT description.

In the same vein, reference \cite{Cribiori:2023ihv} proposes that 5$d$ scale-separated AdS vacua are forbidden and speculates this to hold for arbitrary $d$. The authors make use of the species scale $\Lambda_{\rm sp}$ that has been proposed to define the cutoff of any effective theory coupled to a number $N_{\rm sp}$ of light species \cite{Arkani-Hamed:2005zuc,Distler:2005hi,Dimopoulos:2005ac,Dvali:2007hz,Dvali:2007wp,Dvali:2010vm}:
\begin{equation}
\Lambda_{\rm sp}=\frac{\Mp}{N_{\rm sp}^{\frac{1}{d-2}}}\, .
\end{equation}
The number of species in the bulk is assumed to be equal to the one in the CFT dual which can be estimated with the central charge $c$ like $N_{\rm sp}\sim c$. In 5$d$ AdS, the potential is also related to the central charge through $\Mp^{-2}|V|\sim c^{-\frac{2}{3}}$ from which one can deduce
\begin{equation}
\Lambda_{\rm sp}\sim \sqrt{|V|}\, .
\end{equation}
Under the assumption the the specie scale cannot be decoupled from the KK scale, the formula above forbids scale separation for 5$d$ AdS and one can naively expect a similar reasoning to hold in arbitrary dimension.

\subsubsection{Integer conformal dimensions}
\label{sec:integer}

In this section we mention a striking feature of the CFT duals to some scale-separated vacua and in particular to the original supersymmetric  DGKT ones, namely the fact that the conformal dimensions of the low-lying scalars take integer values. This has first been uncovered in \cite{Conlon:2021cjk} for the vanilla toroidal and supersymmetric DGKT setup and shown to surprisingly also hold for arbitrary Calabi--Yau geometries\footnote{See also \cite{Ning:2022zqx} for similar investigations in M-theory.} in \cite{Apers:2022tfm}, as expected from the universal behaviour of the mass spectrum uncovered in \cite{Marchesano:2019hfb}. Generic features of AdS vacua with integer dual conformal dimensions have been explored in \cite{Apers:2022vfp} while a systematic study of the dual scaling dimensions of all the AdS branches found in \cite{Marchesano:2019hfb} have been performed in \cite{Quirant:2022fpn}. As an example, if one uses \eqref{eq:Delta} and inserts the AdS radius \eqref{eq:Rads} as well as the mass spectrum displayed in \eqref{universal_spectrum}, one finds the following conformal dimensions:
\begin{equation}
\Delta_{\rm s}=\left(6,10,6,2\right)\, ,\quad\begin{cases}
\text{SUSY: }  &\Delta_{\rm a}=(5,11,5,3)\\
\text{non-SUSY: }  &\Delta_{\rm a}=(2,8,8,3)\
\end{cases}\, .
\end{equation}

It was shown in \cite{Quirant:2022fpn} that the conformal dimensions can cease to be integers in some non-supersymmetric families of solutions even though the non-perturbative stability and hence the existence of a proper CFT dual in this case is dubious as we have seen in sect.~\ref{sec:unstab}. About non-supersymmetric solutions, \cite{Apers:2022zjx} also found integer conformal dimensions for a large set of would-be CFT's dual to scale-separated 4$d$ AdS vacua. However this property has been shown to fail for the would-be two-dimensional CFT duals of the 3$d$ scale-separated AdS vacua presented in sect.~\ref{sec:3d}. The question of the origin of these integer conformal dimensions has been investigated in \cite{Plauschinn:2022ztd} by analysing type IIB mirror duals of DGKT setups. As a result, it seems that integer dimensions only arise when the type IIB superpotential nicely splits into two parts, one depending only on the Kähler moduli and the other on the complex structure fields, which is the case for the DGKT mirror; and also when one neglects perturbative corrections to the large complex structure regime.

\subsubsection{Reconstructing large dimensions}

The question of how many large dimensions (i.e. with a size parametrically of the order of the AdS$_d$ radius) a holographic CFT in $d-1$ dimensions can reconstruct, deduced purely from CFT data, has been investigated in \cite{Alday:2019qrf}. The idea is that one-loop amplitudes in the bulk are sensitive to what is running inside and in particular the KK modes associated with large internal dimensions. These one-loop amplitudes, whose structure is constrained by unitarity, are evaluated on the CFT side as scalar four-point correlation functions involving a primary operator $\phi$:
\begin{equation}
\langle\phi(0)\phi(z,\bar z)\phi(1)\phi(\infty)\rangle\, .
\end{equation}
It is shown that the most relevant contribution to the one-loop amplitude comes from the exchange of double-trace operators $[\mathcal{O}\mathcal{O}]_{k,l}$ indexed by $k,l\in\mathbb{N}$. Up to irrelevant terms, the one-loop amplitude is then expanded in conformal blocks as a sum over $k$ and $l$ with coefficients $\beta_{k,l}^{\rm 1-loop}$. Crucially, these coefficients are shown on the one hand to depend on the density of single-trace (ST) operators $\rho_{\rm ST}(\Delta_\mathcal{O})$ with a closed formula and, on the other hand, their asymptotic behaviour at large $k$ can be uncovered. Schematically, one has that
\begin{equation}
\beta_{k,l}^{\rm 1-loop}\propto \sum_\mathcal{O}\rho_{\rm ST}(\Delta_\mathcal{O})\#_{\mathcal{O},k,l}\qquad\text{ and }\qquad\beta_{k\gg 1,l}^{\rm 1-loop}\sim k^{D+d-4}\, ,
\end{equation}
where $D$ counts the number of large dimensions, which includes the $d$ dimensions of AdS and the additional compact ones that have a size of the same order. Knowledge of the single-trace density of operators and how it scales when the conformal dimension $\Delta_{\mathcal{O}}$ is large (translated into a scaling with $k$) can then be used to infer the number $D$ of large dimensions. In a given setup, if $D=d$ then one could conclude that scale separation is achieved while having $D>d$ would signal only a partial separation or none.

To connect and extend with \cite{Polchinski:2009ch} the authors propose a conjecture for what are the sufficient requirements to get a scale-separated AdS. The absence of R-symmetry is not identified as the crucial ingredient in addition to the large gap to higher-spin operators but it is more generally the absence of global symmetries that matters. Of course stating these conditions is one thing and coming up with CFT candidates, even for the more modest case of partial scale separation is another challenge.

\subsubsection{Constraints from logarithmic corrections}

In \cite{Bobev:2023dwx} the authors use heat kernel methods to evaluate quantum corrections to gravitational partition functions and black hole entropy in various setups. Of particular interest are logarithmic corrections which are motivated (from the CFT side) to be independent on the continuous parameters of the gravitational backgrounds. This contrasts with the computations done by the authors on the gravity side that generically produce non-constant logarithmic corrections in that sense. To reconcile the two pictures, the authors conjecture that the part of the spectrum contributing to the logarithmic corrections should be very specific for the contributions to cancel among each other and produce a consistent holographic gravitational theory. When applied to scale-separated vacua, this would impose sharp constraints on the light spectrum and it would be interesting to check these constraints in the various scale-separated scenarios presented in this review.

\section{Conclusion}
\label{sec:conclusion}

Parametric scale separation, i.e. the parametric decoupling of the scales associated with both the internal and compact spaces is a phenomenologically desired property to make sense of low-dimensional effective field theories in string theory. Of course the ultimate goal it to connect with our universe and the study of scale separation in AdS is then just a modest step towards building realistic vacua for which an uplifting to dS is required. But even in this modest setup and despite decades of progress on the subject, it is still not clear and conclusive if one can achieve such a separation of scale in the context of string theory.

On the one hand, a bunch of bottom-up constructions, from a low-dimensional EFT point of view and in the smeared approximation have been proposed to generate infinite families of scale separated vacua in a controlled regime where the volumes are large and the coupling is weak. These constructions have faced many criticisms regarding their 10$d$ and M-theory completions but in light of the most recent works dedicated to these questions, they have passed a series of non-trivial tests to address them. Despite these optimistic results, one can of course not conclude that parametric scale separation is fully consistent with string theory since one can always worry that such progress, both regarding the issue of source localisation and the construction of an M-theory lift, rely only on first-order perturbative analyses and that no fully exact solution is known so far.

On the other hand, different approaches in the context of the Swampland Program and from holography have suggested pessimistic no-go's regarding the possibility to have a hierarchy of scales between spacetime and the internal space. Some of the most established swampland conjectures have been put to use to demonstrate that too many supersymmetries are incompatible with scale separation while other stronger ones (in particular the Strong AdS Distance Conjecture and the AdS Instability Conjecture) just straightforwardly put scale-separated vacua in the Swampland. The lack of an holographic description of scale-separated bulk spacetimes is also suspicious and may signal that all scale-separated constructions are secretly unstable. Conjectures about some CFT properties have even been raised to argue that parametric scale separation is impossible. Of course the absence so far of an explicit CFT dual to a scale-separated gravitational theory is not a proof that it does not exist, and the strongest conjectures that ban scale separation may not be true in full generality unless they are weakened in some various ways. Although more pessimistic than the bottom-up approach, these Swampland and holographic perspectives have thus also been unable to reach fully conclusive answers so far.

Scale separation is caught between these two approaches to determine where the borderline of the Landscape should be and if it includes this property or not. More work is needed and the recent headway made on both sides is promising for arriving at robust conclusions in the future.

\bigskip
\bigskip

\centerline{\bf  Acknowledgments}

\vspace*{.5cm}

We warmly thank Arthur Hebecker, Fernando Marchesano and Thomas Van Riet for feedbacks, as well as Björn Friedrich, Ruben Küspert, Miguel Montero and David Prieto for useful discussions.

\bibliography{biblio}

\end{document}